\documentclass[10pt]{article}%

\usepackage[a4paper,top=2.5cm,left=2cm,bottom=2.5cm,right=2cm]{geometry}%b5paper
\usepackage{amsmath}
\usepackage{amssymb}
\usepackage{graphicx}
\usepackage{hyperref} %highlight hyperlink
\usepackage[authoryear,sort&compress]{natbib} %\citep 
\usepackage{subfigure}  %= use for side-by-side figures

\newcommand{\sci}{Science}
\newcommand{\aap}{Astronomy and Astrophysics}
\usepackage{lineno}

\begin{document}
\title{Stability of cycle in samplogram and spurious cycles in solar activity
}

\author{Kim Chol-jun\\%
\small \textit{Faculty of Physics, Kim Il Sung University, DPR Korea}
\\%\and 
\centering{ \small email address: cj.kim@ryongnamsan.edu.kp}
}

\maketitle
%\linenumbers
\begin{abstract}
The spectral analysis with stochastic significance cannot distinguish the spurious cycles effectively, because a non-stationary signal can make a significant peak in spectrum. I show that the random separation between the grand extremes such as grand maxima and minima in solar activity can make a spurious but significant peak in spectrum. And it is possible to pick even the weak stable cycle by applying an averaging down-sampling and the differencing to a random signal. This is because both operations variously change the height of peak in spectrum and the spurious cycle is in turn unstable in both operations. I introduce a samplogram showing these operations and stability intuitively.
\end{abstract}

%\paragraph{keywords:}Sun: activity; methods: data analysis;

\section{Introduction}\label{sec:intro}
The cycles in solar activity have long been discussed. In mid-19th century Schwabe and Wolf had discovered the well-known 11-yr solar cycle. Later, the Gleissberg cycle ($\sim$80 yr) was proposed, which, however, has dimly ($60 \sim 120$ yr) identified in the spectrum. 

The reconstruction of the past solar activity makes it possible to study the long-term solar cycles. The reconstruction from cosmogenic radionuclides such as $^{14}$C and $^{10}$Be  \citep{Eddy1976, Solanki2004, Usoskin2008, Steinhilber2012, Usoskin2016, Beer2018, Wu2018, Chol-jun2020a} have implied the various solar cycles: The Gleissberg (88-yr), Suess/de Vries (208-yr), Hallstatt ($\sim$2400-yr), Eddy ($\sim$1000-yr), unnamed $\sim$350-yr, $\sim$500 and $\sim$710-yr cycles. However, the identification of those cycles are not certain yet. Only the Hallstatt modulation of the Suess/de Vries cycle may be a mere cross-related argument of their existence.

The cycles have been studied in the historical records of sunspots and aurorae.
\citet{Xu1990} analyzed his catalog of historical naked-eye sunspot observations and reported that beside 10.86-yr period 212-yr period is most significant cycle of solar activity. The Suess/de Vries cycle in the historical records were reported by other authors \citep{Ma2009}. \citet{Chol-jun2020a} analyzed Korean medieval sunspot records and found the trace of Suess/de Vries cycle. \citet{Ogurtsov2002} analyzed the data on ancient sunspot observations and found the evidence of the century-type cycle of Gleissberg in a wide frequency band with a double structure consisting of 50-80 yr and 90-140 yr periodicities and the Suess/de Vries cycle with a period of 170-260 yr.  In the records of historical aurorae, the 88-yr and, specifically, 130-yr cycles have been reported in \citet{Attolini1988}. Although the historical records have uneven spacing and vague magnitude, the cycles inferred from the strongly noisy records should have some robustness.

\citet{Cameron2019} have shown that fluctuations in the power spectrum of sunspot numbers (SSN) reconstructed from the cosmogenic isotopes are consistent with realization noise with no intrinsic periodicities except that of the basic 11/22-yr cycle. They simulated significant spurious peaks of the Gleissberg and Suess/de Vries cycles from model for a noisy and weakly nonlinear limit cycle. Their report is at odds with the previous analyses for the historical and reconstruction datasets. We can give a question: can the small randomness be accumulated to make a significant spurious peak in spectrum? Did this happen to the past solar activity? We want to have more intuitive and phenomenological explanation for the ``significant spurious'' peak in the random process and the solar activity. As we will see later in text, the non-iterative and random separation between the grand extremes (maxima and minima) can make such a ``significant spurious'' peak in spectrum. And this shows a drawback of the ordinary spectral analyses (the Fourier and power spectra) in distinguishing the true cycles from the spurious cycles.

There are several advanced methods of spectral analysis. The Lomb-Scargle (LS) periodogram is a typical one. This method has an advantage to be more flexible in the time and frequency domain in comparison with the Fourier method so that this can analyze the unevenly-spaced time series and give the power at any frequency, though the independent frequencies are limited \citep{Scargle1989}. This periodgram also includes a parameter to distinguish the true peak from the random peak. This is the false alarm probability (FAP) which shows that a significant peak in LS periodogram happens hardly due to the random signal. But the low peak can be made easier from the random signal. If the amplitude of a true cycle is small, this also should make the low peak. So we are probable to miss a low true peak and pick a significant spurious peak. 

Where comes the drawback of the existing spectral analyses from? We consider the formation of peak of true cycle in spectrum. As \citet{Scargle1989} indicated, the peak forms at which the frequency is consistent with one of the cycle. Then the height of peak reflects the amplitude of cycle. Thus, the peak represents both the frequency consistency and the cycle amplitude. The aforementioned FAP is based on the peak height, and this cannot determine the cycle amplitude and the fact of whether the frequency corresponds to the true cycle, simultaneously. For example of an intermittent cycle, the persistence  and amplitude of the cycle are reflected together on the peak height. This also becomes a drawback of the method based on the stochastic significance evaluation.

In this paper, we discuss the stability of cycle. This does not imply only the time stability in the above example of intermittent cycle. The stability of cycle means how stably the peak of cycle maintains in any cycle-keeping operation on the signal or its time series. In the previous work \citep{Chol-jun2020a} we proposed the samplogram method which shows how the peaks change in variation of the sampling interval and via the differencing. If we see an object with a movement, we can recognize the object better. This idea has been included also in the other works. \citet{Beutler1966} tried to give a robust spectral analysis, using the point process for random sampling where the points in time series are chosen randomly but with some distribution rule. What is a cycle-keeping operation? This operation is that keeps any cycle when applied to the signal. A variation of the sampling interval is an instance, provided that the sampling rate is higher than the Nyquist rate. The stationary point process for random sampling in \citet{Beutler1966} can be regarded as cycle-keeping. The differencing is also cycle-keeping. The cycle-keeping operation is similar to the symmetric transformation in the field theory. 

Then what properties are desired for the cycle-keeping operation? First, to avoid a drawback of aforementioned stochastic significance method, we require that the amplitude of different cycles should change differently so that the frequency consistency could be separated from the amplitude variation. It would be ideal that the amplitude of different cycles change independently. At least, the amplitude of different cycles should not change with the same scaling. Secondly, it is desired that the operation parameter changes successively (or continuously), because we can observe the variation of peak as a successive (or continuous) process. 

In this context, we see the sampling and differencing (SnD) stability with a samplogram in this paper. We show that even a negligible true cycle appears to be stable in the samplogram. The samplogram can also intuitively show the abnormal aliasing effect related to the higher-rate signal than the sampling rate. 

\section{The power spectrum of a multi-cyclic sinusoidal signal}\label{sec:ps}
In this section we see the variation of peak for the sinusoidal signal when the sampling interval changes.  

\subsection{The leading-order approximation for the power of sinusoidal signal in an averaging down-sampling}\label{subsec:apprps}
\newcounter{eqswap}
We start with a multi-cyclic sinusoidal signal: 
\begin{equation}\label{eq:startsignal}
u(t)=\sum _j U_j \cos \left(\frac{2\pi t}{T_j}+\phi _j\right),
\end{equation}
where $U_j, T_j \text{ and } \phi _j$ are the amplitude, period and phase of the $j$-th mode of the signal, respectively. A time series sampled from this signal with a time-step $\Delta t$ is expressed as
\begin{linenomath}\begin{align}\label{eq:oritmsrmlti}
u(r)&= \sum _j U_j \cos \left(\frac{2 \pi r \Delta t}{T_j}+\phi_j\right) %\nonumber \\&
=\frac{1}{2}\sum _j  U_j \left[\exp(i \phi_j) \exp \left(\frac{2\pi  i r \Delta t}{T_j}\right)+\exp(-i \phi_j) \exp \left(-\frac{2\pi i r \Delta t}{T_j}\right)\right],
\end{align}\end{linenomath}
where $r$ is the index of point in the time series and runs from 1 to $N$ that is the length of the time series. Also, $t\approx r\Delta t$.  

Apply a down-sampling to the time series, which means the sampling interval getting greater or a coarser sampling. The down sampling can have two approaches: hopping and averaging. A hopping down-sampling takes points of the series after every several steps (e.g. $m$ steps) to make a new series, which causes the loss of information because some data points are missed in the down-sampled series. However, the data points of the down-sampled series are already included in the original series. When the sampling interval increasing in this down-sampling, the total number of points is reduced by the same factor as the sampling interval increases. This approach is used in signal acceptance of machines with cares of low acceptance frequency. The hopping down-sampled series is expressed as
\begin{linenomath}\begin{align} \label{eq:umkexpmulti}
u_{m,k}(r)=\frac{1}{2}\sum _j  U_j \bigg[\exp(i \phi_j) \exp \left(\frac{\displaystyle 2\pi  i  \Delta t(k+m (r-1))}{T_j}\right) %\nonumber \\
+\exp(-i \phi_j) \exp \left(-\frac{\displaystyle 2\pi  i \Delta t(k+m (r-1))}{T_j}\right)\bigg],
\end{align}\end{linenomath}
where $m$ is the sampling step and $k$ is an integer between 1 and $m$, that is, we can select one series among $m$ down-sampled series. $r=1 \sim n$ is the index of point in the down-sampled series. The length of the down-sampled series is approximated to be
\begin{linenomath}\begin{equation}\label{eq:n}
n=\texttt{IntegerPart}(N/m),
\end{equation}\end{linenomath}
that is, if a perfect factorization $N = m n$ is impossible, then integer part of $N/m$ is taken as the length of the down-sampled series, which means that in down-sampling the series is truncated. However, this approach will simplify analysis.

On the other hand, an averaging down-sampling is to average every $m$ points and take it as a point of the new down-sampled series. The averaging down-sampled series is expressed as
\begin{linenomath}\begin{align}\label{eq:umbarexp}
u_{\bar{m}}(r)&=\frac{1}{m} \sum _{k=1}^m u_{m,k}(r)  \\
&=\frac{1}{2m}\sum _j U_j \bigg[ \exp(i \phi_j) \exp \left(\frac{\displaystyle 2\pi  i (r-1)m \Delta t}{T_j}\right)\exp \left(\frac{\displaystyle 2\pi  i \Delta t}{T_j}\right) \frac{1-\exp \left( \frac{\displaystyle 2\pi  i m \Delta t}{T_j}\right) }{1-\exp \left( \frac{\displaystyle 2\pi i \Delta t}{T_j}\right)} \nonumber \\
&\quad+ \exp(-i \phi_j) \exp \left(-\frac{\displaystyle 2\pi  i (r-1)m \Delta t}{T_j}\right)\exp \left(-\frac{\displaystyle 2\pi  i \Delta t}{T_j}\right) \frac{1-\exp \left( -\frac{\displaystyle 2\pi  i m \Delta t}{T_j}\right) }{1-\exp \left( -\frac{\displaystyle 2\pi i \Delta t}{T_j}\right)} \bigg], \nonumber
\end{align}\end{linenomath}
which can be derived from Eq.~\eqref{eq:umkexpmulti} by using formula for the geometric series:
\begin{linenomath}\begin{align}\label{eq:geoser}
\sum _{k=1}^m \exp \left(\frac{\displaystyle 2\pi  i k \Delta t}{T}\right)=\exp \left(\frac{\displaystyle 2\pi  i \Delta t}{T}\right) \frac{1-\exp \left( \frac{\displaystyle 2\pi  i m \Delta t}{T}\right) }{1-\exp \left( \frac{\displaystyle 2\pi i \Delta t}{T}\right)}.
\end{align}\end{linenomath}
In the hopping down-sampling the data points are inherited intact to the down-sampled series. This gives a risk to inherit the spurious cycle. However, in the averaging down-sampling, the new data points are forged by averaging and, in that way, the noise is suppressed and the risk that the spurious cycle is inherited intact to the down-sampled series lowers. From now on, in this paper, we will mainly consider the averaging down-sampling.

The power spectrum is obtained by the Fourier transform of autocorrelation of the signal. If two sampled signals $x(r)$ and $y(r)$ are given, the correlation between them is defined as a normalized covariance:
\begin{linenomath}\begin{equation}\label{eq:cor1}
\texttt{Cor}(x,y)=\frac{\texttt{Cov}(x,y)}{\sqrt{\texttt{Var}(x)\texttt{Var}(y)}}.
\end{equation}\end{linenomath}
$\texttt{Cov}(x,y)$ stands for the covariance between $x(r)$ and $y(r)$ of length $N$:
\begin{linenomath}\begin{align}\label{eq:cov1}
\texttt{Cov}(x,y)=\frac{1}{N-1}\sum _{r=1}^N (x(r)-\bar{x})(y(r)-\bar{y})^*,
\end{align}\end{linenomath}
where the asterisk means complex conjugate. If we deal only with real signals, the conjugate symbol disappears. $\bar{x}$ and $\bar{y}$ stand for the mean values of $x(r)$ and $y(r)$: $\bar{x}=\frac{1}{N}{\sum _{r=1}^N x(r)}\text{ and } \bar{y}=\frac{1}{N}{\sum _{r=1}^N y(r)}$.
The variances of $x(r)$ and $y(r)$ are defined in a similar way as the covariance:
\begin{linenomath}\begin{align}\label{eq:varx1}
\texttt{Var}(x)=\frac{1}{N-1}\sum_{r=1}^N (x(r)-\bar{x})(x(r)-\bar{x})^*, \quad 
\texttt{Var}(y)=\frac{1}{N-1}\sum_{r=1}^N (y(r)-\bar{y})(y(r)-\bar{y})^* .
\end{align}\end{linenomath}

The autocorrelation of a sampled signal $x(r)$ is defined in the same way as the correlation (Eq.~\ref{eq:cor1}).
\begin{equation}\label{eq:ac1}
AC(x,g)=\frac{\texttt{Cov}(x(r),x(r+g))}{\sqrt{\texttt{Var}(x(r))\texttt{Var}(x(r+g))}},
\end{equation}
where $g$ is shifting steps (or lag), that is, the autocorrelation of signal can be the correlation between a subseries $x(r)$ starting at the first point of the series and some (e.g. $g$) steps shifted (or lagged) subseries $x(r+g)$ ending at the last point of the series. If the length of the time series is fixed, the lengths of both shifted series must be truncated by the shifting steps. 

To evaluate the power spectrum or autocorrelation (AC) periodogram we use the discrete Fourier transform (DFT). The DFT $v(s)$ of a time series $u(r)$ with time-step $\Delta t$ and length $N$ is defined as 
\begin{linenomath}\begin{align}\label{eq:dtf}
v(s)=\frac{1}{\sqrt{N}}\sum _{r=1}^N u(r) \exp \left( \frac{2 \pi  i (r-1) (s-1)}{N}\right),
\end{align}\end{linenomath}
where $s$ is the index of point of the series $v(s)$ and stands for the frequency index. The frequency and period corresponding to the $s$-th point of the series $v(s)$ are evaluated as 
\begin{equation}\label{eq:DFTper}
f=s\Delta f \qquad \text{and}\qquad  T=\frac{N \Delta t}{s-1}.
\end{equation}
$s$ runs from 2 to $N$ and the frequency step $\Delta f$ has relation with the time step $\Delta t$ as $\Delta f \Delta t=\frac{1}{N}$. For simplicity, take following replacements: 
\begin{linenomath}\begin{align}
s-1&=q,  \label{eq:q}\\
\frac{\Delta t}{T_j}=\frac{q_{0j}}{N}&\approx\frac{q_{0j}}{mn}. \label{eq:T_j}
\end{align}\end{linenomath}
The frequency index $s$ is replaced with $q=s-1$ and the period $T_j$ of the $j$-th mode is replaced with frequency $q_{0j}$. Remember that in the averaging down-sampling the whole length of the series $N$ is truncated into a reducible product $mn$ where $m$ is averaging step and $n=\texttt{IntegerPart}(N/m)$ is the length of the down-sampled series(Eq.~\ref{eq:n}).

\begin{figure}
\begin{center}
\subfigure[]{
\includegraphics[scale=0.4]{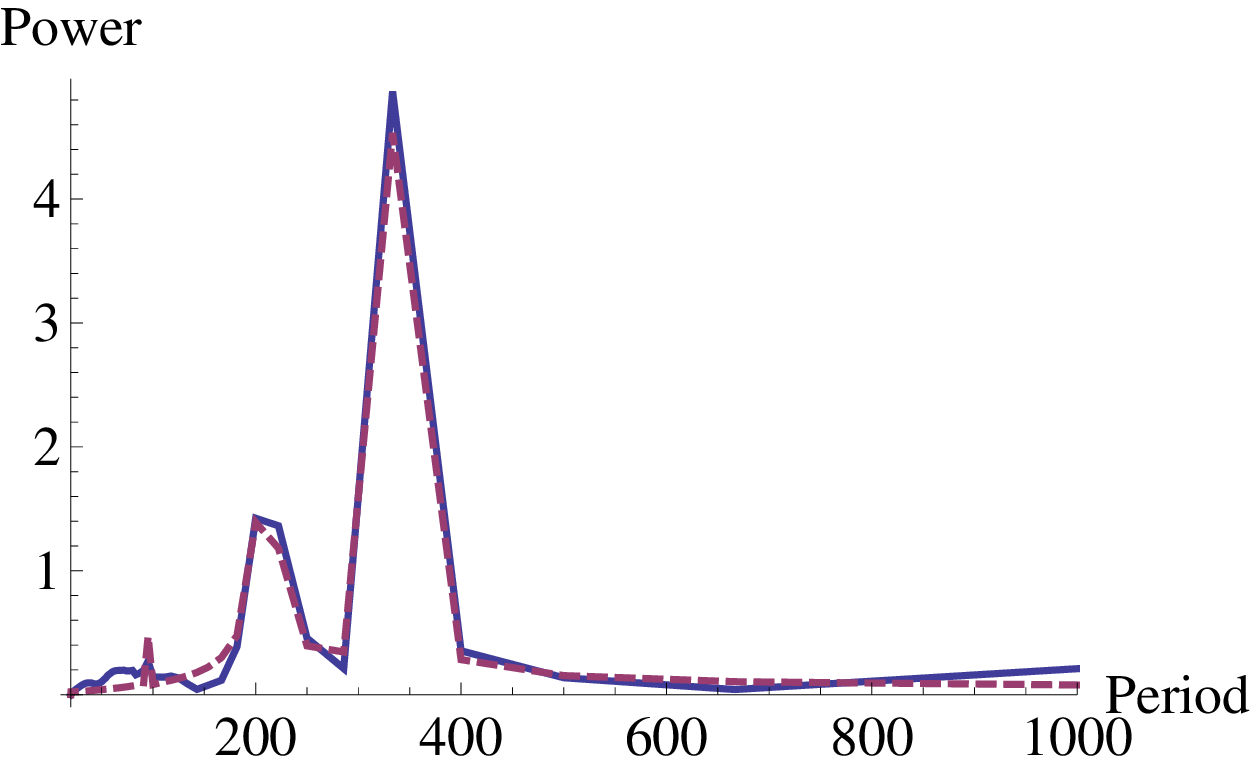}}
\subfigure[]{
\includegraphics[scale=0.4]{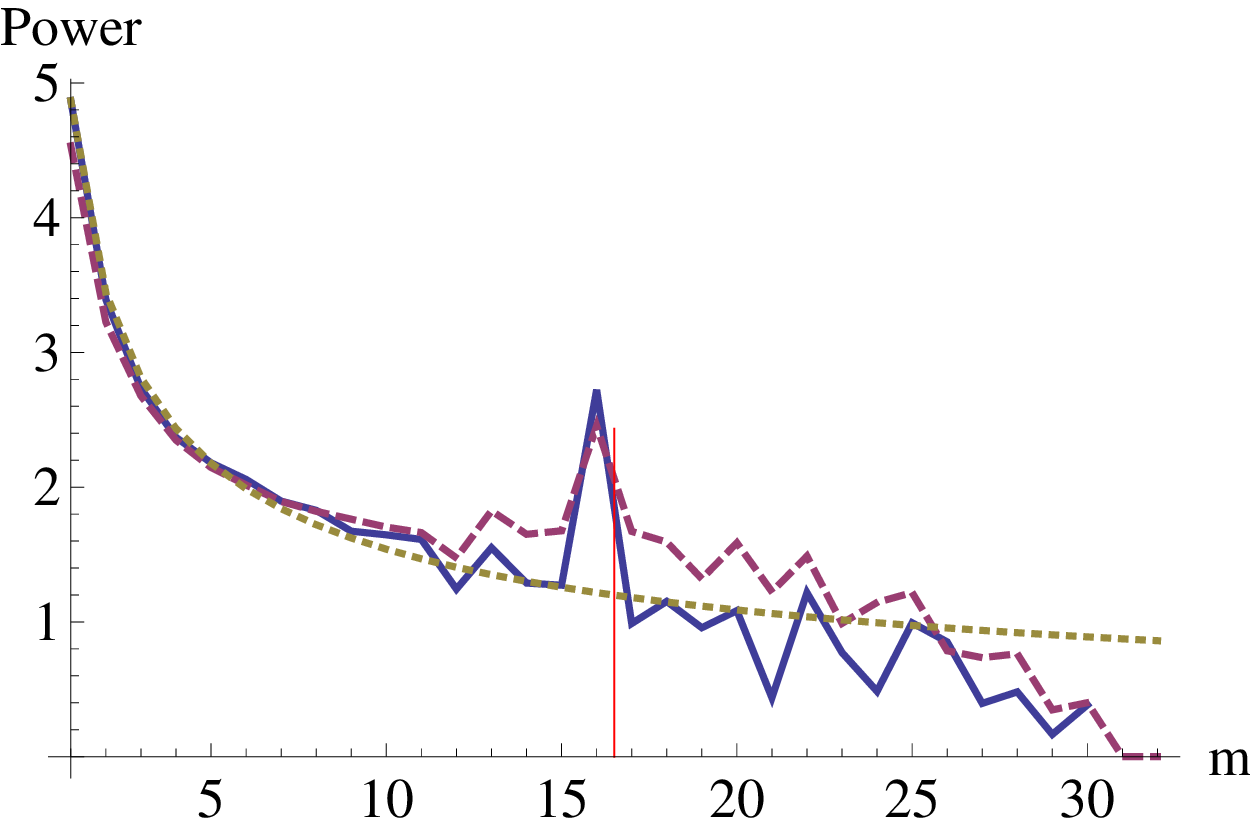}}
\caption{\label{fig:PS0} \small The power spectrum and variation of the power in averaging down-sampling of a composite signal of $U_1=1, U_2=2, U_3=3, T_1=95, T_2=210, T_3=330, \phi_1=1.5, \phi_2=1, \phi_3=2.4, \Delta t=10 \text{ and }N=200$. (a) The power spectrum at sampling step $m=1$. (b) Variation of power of the $j=3$-rd mode when increasing the sampling step $m$. The solid line shows the numerical evaluation and dashed line does the approximate one. The dotted line in (b) means the degradation proportional to $\frac{1}{\sqrt{m}}$. The red vertical bar means the Nyquist sampling step for the 3-rd mode: $m=\frac{T_3}{2\Delta t}=16.5$. At the Nyquist sampling the sampling peak appears.}
\end{center}
\end{figure}

Now, it is possible to evaluate the power or AC periodogram of the down-sampled time series $u_{\bar{m}}(r)$ of length $n$ by Eq.~\ref{eq:dtf}:
\begin{linenomath}\begin{align}  \label{eq:ACPer}
PSm(s)=\frac{1}{\sqrt{n}}\sum _{g=1}^n ACm(g) \exp \left( \frac{2 \pi  i (g-1) (s-1)}{n}\right),
\end{align}\end{linenomath}
where $ACm(g)$ is the autocorrelation series of the averaging down-sampled time series $u_{\bar{m}}(r)$ and the shifting step $g$ is used as a time variable for the Fourier transformation. Do not confuse $g$ with the frequency index $q$. Taking care of Eq.~\eqref{eq:q}, a long algebraic calculation with numerical comparisons gives us the expression for AC periodgram at the leading order
\newcounter{eqpsfinal}
\setcounter{eqpsfinal}{\value{equation}}
\begin{linenomath}\begin{align}\label{eq:psfinal}
PSm&(q)=\frac{1}{\sqrt{n}}\sum _{g=1}^n ACm(g) \exp \left( \frac{2 \pi  i (g-1) q}{n}\right) \nonumber \\
&\approx \frac{1}{2 \sqrt{n}}\sum _{g=1}^n \frac{\sum_j U_j^2 A_j}{\sum_k U_k^2 A_k} \left[ \exp \left(\frac{2 \pi  i g q_{0j}}{n}\right)+\exp \left(-\frac{2 \pi  i g q_{0j}}{n}\right)\right] \exp \left( \frac{2 \pi  i (g-1) q}{n}\right) \nonumber \\
&=\frac{1}{2\sqrt{n}\sum_k U_k^2 A_k}\sum_j U_j^2 A_j  %\nonumber\\& \qquad \times
\left[\exp\left(\frac{2 \pi  i q_{0j}}{n}\right) \frac{1-\exp\left(2 \pi  i(q+q_{0j})\right)}{1-\exp\left(\frac{2 \pi  i(q+q_{0j})}{n}\right)}+\exp\left(-\frac{2 \pi  i q_{0j}}{n}\right) \frac{1-\exp\left(2 \pi  i(q-q_{0j})\right)}{1-\exp\left(\frac{2 \pi  i(q-q_{0j})}{n}\right)} \right],
\end{align}\end{linenomath}
where 
\begin{linenomath}\begin{align} \label{eq:Ajq}
A_j=\frac{1-\exp \left(\displaystyle \frac{ 2 \pi  i q_{0j}}{n}\right) }{1-\exp \left(\displaystyle \frac{ 2 \pi  i q_{0j}}{mn}\right)} \;
\frac{1-\exp \left(\displaystyle -\frac{ 2 \pi  i q_{0j}}{n}\right) }{1-\exp \left(\displaystyle -\frac{ 2 \pi  i q_{0j}}{mn}\right)}.
\end{align}\end{linenomath}
Using Eq.~\eqref{eq:psfinal}, we can simulate the power spectrum of the signal and variation of the power in the averaging down-sampling. Fig.\ref{fig:PS0} shows the numerical and approximate evaluations of power spectrum and the variation of the power in the averaging down-sampling (the numerical one is evaluated by the default functions in language \textsl{Mathematica} and the approximate one is by Eq.~\ref{eq:psfinal}). The numerical and approximate evaluations show a good agreement.

In Eq.~\eqref{eq:psfinal}, the quantity in square brackets reflects the frequency dependency and gives some ``peaks":
\begin{linenomath}\begin{align}\label{eq:pscore}
PSmcore(q)=\left[\exp\left(\frac{2 \pi  i q_{0j}}{n}\right) \frac{1-\exp\left(2 \pi  i(q+q_{0j})\right)}{1-\exp\left(\frac{2 \pi  i(q+q_{0j})}{n}\right)}+\exp\left(-\frac{2 \pi  i q_{0j}}{n}\right) \frac{1-\exp\left(2 \pi  i(q-q_{0j})\right)}{1-\exp\left(\frac{2 \pi  i(q-q_{0j})}{n}\right)} \right],
\end{align}\end{linenomath}
where there appear two "singularities" at $q=q_{0j}$ and $q+q_{0j}=n$. In fact, they are not the real singularities. In approaching $q\rightarrow q_{0j}$, we find that
\begin{linenomath}\begin{align} \label{eq:sing}
\lim_{q\to q_{0j}} \frac{1-\exp\left(2 \pi  i(q-q_{0j})\right)}{1-\exp\left(\frac{2 \pi  i(q-q_{0j})}{n}\right)}=n.
\end{align}\end{linenomath}
This corresponds to the maximum of the left hand side. So the both peaks appear: the former ``singularity" in Eq.~\eqref{eq:pscore} makes \textit{a true peak} at  frequency $q=q_{0j}$ in spectrum and the latter does \textit{an aliasing peak} at frequency $q=n-q_{0j}$. The frequency domain of the Fourier spectrum is given in $0 \leq f \leq f_s$, where $f_s=\frac{1}{\Delta t}$ is a sampling frequency. A region of $0 \leq f \leq \frac{f_s}{2}$ is a true frequency region and a region of $\frac{f_s}{2} \leq f \leq f_s$ is an aliasing frequency region. According to the sampling theorem, only modes of frequency $f<\frac{f_s}{2}$, i.e. frequency less than the Nyquist sampling frequency $f_N=\frac{f_s}{2}$, are effectively evaluated. The aliasing peak at $q=n-q_{0j}$ in the aliasing region $\frac{f_s}{2} \leq f \leq f_s$ is no more than a reflection of the true peak at $q=q_{0j}$ in the true region, which is obvious from a symmetry of the above expressions. Ordinarily, the aliasing region (and aliasing peak) is neglected in the power or Fourier spectra. Actually, the frequency $q_{0j}$ at which the true peak occurs does not depend on the sampling interval $m \Delta t$ or the length $n$ of the down-sampled series. This says that when increasing the sampling interval $m$ the true peak keeps at the same frequency $q=q_{0j}$. This is just \textit{a down-sampling (or sampling) stability of cycle}. Such endured peaks when increasing sampling interval form a track of peaks like a ridge, which we will see later. 

In fact, in the DFT, the true peak should happen not exactly at $q=q_{0j}$ but at $q=\texttt{Round}(q_{0j}j)$ and the aliasing peak at $q=\texttt{Round}(n-q_{0j})$ because $q$ is not always integer, where \texttt{Round}() means the rounding value.

If $q=\frac{n}{2}$, the true peak around $q=q_{0j}$ and the aliasing peak around $q=n-q_{0j}$ are overlapped, which makes another type of peak. We call this peak \textit{a sampling peak} of the j-th mode because the overlapping frequency $q=\frac{n}{2}$ corresponds to the Nyquist sampling $m\Delta t=\frac{T}{2}$ (Eq.~\ref{eq:T_j}). In comparison with the sampling peak, the traditional true peak at $m=1$ could be called \textit{a resonance peak}, which represent the cycle in the Fourier and power spectra. In fact, if we consider a case of the Nyquist sampling of the $j$-th mode, it holds
\begin{linenomath}\begin{align}\label{eq:Nyqn}
m_{Nj} \Delta t=\frac{T_j}{2}\qquad \text{and} \qquad n_{Nj}=2q_{0j},
\end{align}\end{linenomath}
from which
\begin{linenomath}\begin{align}\label{eq:psNf}
PSm(q_{0j},n_{Nj})\approx\frac{\sqrt{n_{Nj}}}{2\sum_k U_k^2 A_k} U_j^2 A_j \left[\exp\left(\frac{2 \pi  i q_{0j}}{n_{Nj}}\right)+\exp\left(-\frac{2 \pi  i q_{0j}}{n_{Nj}}\right) \right]
\approx-\frac{\sqrt{n_{Nj}}}{\sum_k U_k^2 A_k} U_j^2 A_j,
\end{align}\end{linenomath}
where $m_{Nj}$ and $n_{Nj}$ are the averaging step and the length of the down-sampled time series $u_{\bar{m}}(r)$ in Eq.~\ref{eq:umbarexp}. Thus, it is obvious that both terms in square brackets in Eq.~\eqref{eq:psNf} have the same phase so they seem not cancel out each other and form the peak at the Nyquist sampling. 

\begin{figure}
\begin{center}
\subfigure[]{
\includegraphics[scale=0.4]{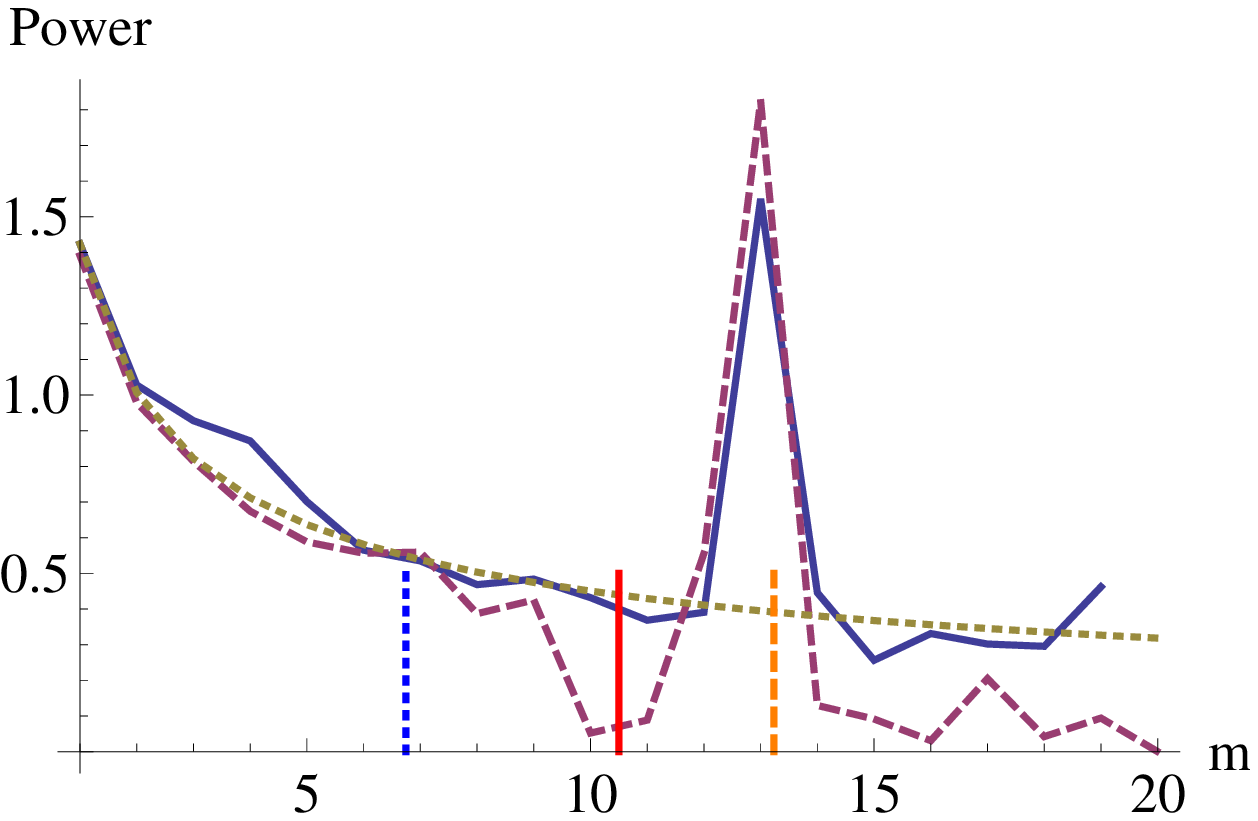}}
\subfigure[]{
\includegraphics[scale=0.4]{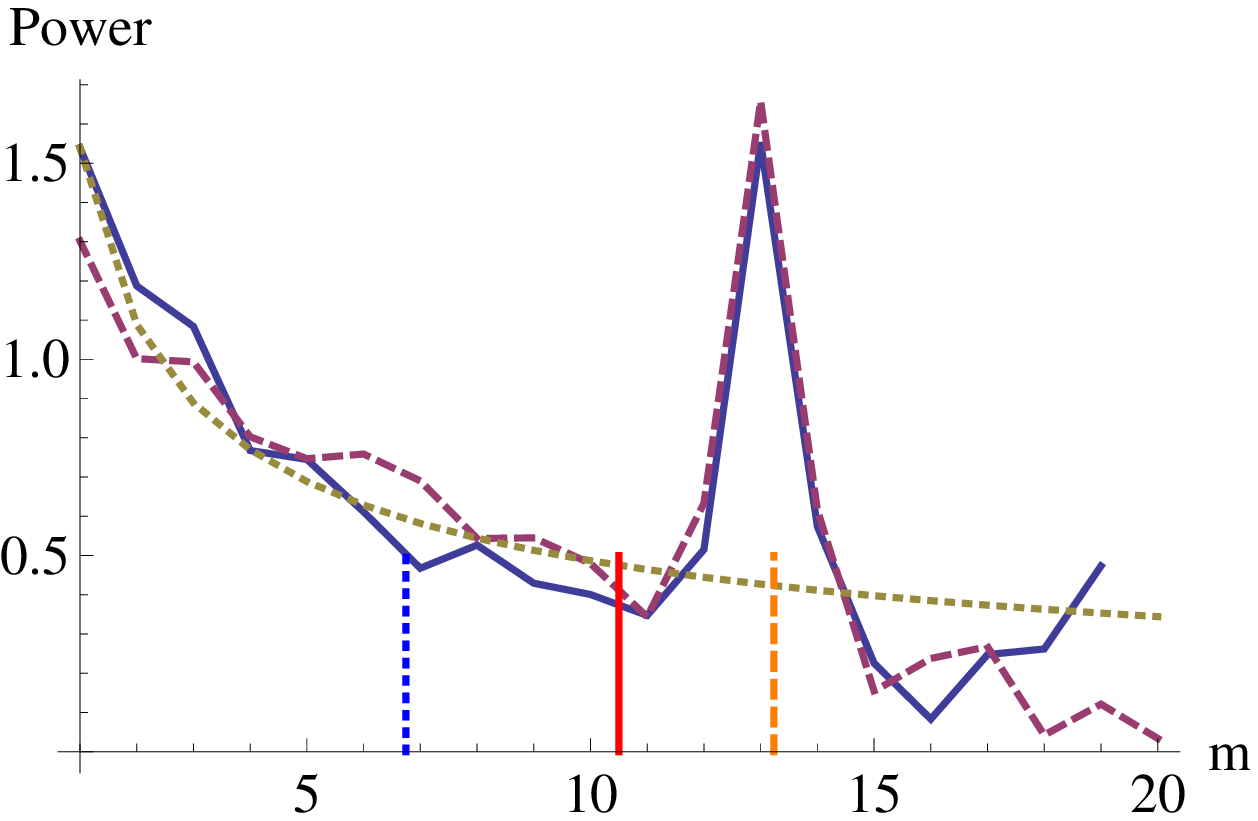}}
\caption{\label{fig:smppkn} \small The variations of the power of the $j=2$-nd mode in Fig.~\ref{fig:PS0}. The length of the Nyquist down-sampled series is (a) $n=20$ and (b) $n=19$. The solid line stands for the numerical evaluation, the dashed line for the approximate evaluation according to Eq.~\eqref{eq:psfinal} and the dotted line for the degradation proportional to $\tfrac{1}{\sqrt{m}}$.  In (a) the approximate evaluation (dashed line) shows a  valley at the Nyquist sampling. In (b) the valley disappears. The red vertical bar means the Nyquist sampling $m=10.5$. The blue dotted vertical bar stands for the nodal peak with the $j$=1-st mode and the orange dashed vertical bar for the nodal peak with the $j$=3-rd mode (see \S~\ref{subsec:cnfgsmp}).}
\end{center}
\end{figure}

However, instead of the sampling peak it is possible to appear a valley. An example is shown in numerical simulation of the power degradation in the down-sampling for the $j=2$-nd mode in Fig.~\ref{fig:PS0}. A valley appears at the Nyquist sampling! (Fig.~\ref{fig:smppkn}(a)) In this case, the power at the Nyquist sampling is evaluated in Eq.~\eqref{eq:psfinal} with $m=10, n=20, q=10 \text{ and }q_{02}=q_{0j}=9.52381$ from $T_2=210$. Here we can find $2q=n$. It seems that the real value of power at $2q=n$ should be related to the appearance of valley through a possibility that the power can become zero. Investigate a value range of the power in Eq.~\eqref{eq:pscore} for the above case. The value of Eq.~\eqref{eq:pscore} approaches to 0 when $q_{0j}$ comes to 9.5238 and 10.4762. In DFT a resonance peak happens at $q=\texttt{Round}(q_{0j})$ instead of $q_{0j}$, so in the case of $q-q_{0j}\approx 0.5$ the power might be zero.

This problem can be avoided if $n$ is odd. For example, we can set $n=2\texttt{IntegerPart}(\frac{n}{2})-1$. Then Eq.~\eqref{eq:pscore} cannot to be real and, simultaneously, we can avoid 0, i.e. the appearance of valley. However, for the odd $n$, the peak at $q_{0j}=q$ becomes lower. Actually, the maximum power lowers by $\sim$2 times than the case of even $n$, which means there disappear a significant sampling peak over the ridge. But the peak becomes broader in frequency domain for the odd $n$ than for the even $n$. In fact, zero points of $q_{0j}$ are 8.0 and 11.0, so the peak broadens 3 times. Thus, we see that the valley at the Nyquist sampling can be avoided by that the length of every down-sampled series should be odd. However, then the peak as well as the valley might disappear (Fig.~\ref{fig:smppkn}b).

Instead of the normalized covariance (Eq.~\ref{eq:ac1}), we can define the autocorrelation as the absolute covariance 
\begin{equation}\label{eq:acab}
AC(x,g)=\texttt{Cov}(x(r),x(r+g)),
\end{equation}
which is popular in many books (cf. \citet{Oppenheim1999}). Then the power of the averaging down-sampled time series $u_{\bar{m}}(r)$ in Eq.~\ref{eq:umbarexp} can be evaluated as
\begin{linenomath}\begin{align}\label{eq:psab}
PSm(q)&\approx \frac{1}{4m^2\sqrt{n}}\sum_j U_j^2 A_j %\times \nonumber\\&\times
\left[\exp\left(\frac{2 \pi  i q_{0j}}{n}\right) \frac{1-\exp\left(2 \pi  i(q+q_{0j})\right)}{1-\exp\left(\frac{2 \pi  i(q+q_{0j})}{n}\right)}+\exp\left(-\frac{2 \pi  i q_{0j}}{n}\right) \frac{1-\exp\left(2 \pi  i(q-q_{0j})\right)}{1-\exp\left(\frac{2 \pi  i(q-q_{0j})}{n}\right)} \right].
\end{align}\end{linenomath}

We can also induce the Fourier transform of the averaging down-sampled time series $u_{\bar{m}}(r)$ more easily:
\begin{linenomath}\begin{align}
v_{\bar{m}}(q)\approx & \sum_j \frac{U_j}{2 m\sqrt{n}}\bigg\{\exp (i\phi_j)\exp\left(\frac{2 \pi i}{m n} q_{0j}\right)\frac{1-\exp\left(\frac{2 \pi i}{n} q_{0j}\right)}{1-\exp\left(\frac{2 \pi i}{m n} q_{0j}\right)}\frac{1-\exp\left[2\pi i(q+q_{0j}) \right]}{1-\exp\left[\frac{2\pi i}{n}(q+q_{0j})\right]} \nonumber\\
&\qquad+\exp (-i\phi_j)\exp\left(-\frac{2 \pi i}{m n} q_{0j}\right)\frac{1-\exp\left(-\frac{2 \pi i}{n} q_{0j}\right)}{1-\exp\left(-\frac{2 \pi i}{m n} q_{0j}\right)}\frac{1-\exp\left[2\pi i(q-q_{0j}) \right]}{1-\exp\left[\frac{2\pi i}{n}(q-q_{0j})\right]}
\bigg\}.
\end{align}\end{linenomath}

It is obvious that the Fourier amplitude and (normalized and absolute) powers have the sampling peak at $q = q_{0j} = \frac{n_{Nj}}{2}$. Furthermore, they are real. So their complex argument could be 0 or $\pm\pi$ and the power of sampling peak should be negative real. However, the DFT violates such an expectation. The Fourier transform is easier than the power to extract the information of cycle such as the amplitude and phase. On the other hand, the power is based on several averaging (in down-sampling and autocorrelation) so that this can depress the random part of the signal more effectively than the Fourier transform.

\subsection{The power variation in the averaging down-sampling}\label{subsec:powvar}
We investigate how the power of a resonance peak changes when increasing sampling interval (i.e. the averaging step) $m$. The variation trend of the power is not simple. At small $m$ and around the true peak, the first term in Eq.~\eqref{eq:pscore} that corresponds to the aliasing peak can be neglected. Also at small $m$ the other modes but the $j$-th mode can not make the resonance peaks around $q\approx q_{0j}$. Thus the power of the resonance peak at $q=\texttt{Round}(q_{0j})$ can be approximately evaluated:
\begin{linenomath}\begin{align}\label{eq:normdec}
\vert PSm(q_{0j},m) \vert \approx \sqrt{n} \frac{U_j^2 A_j }{2\sum_k U_k^2 A_k} \approx \sqrt{\frac{N}{m}} \frac{U_j^2 A_j }{2\sum_k U_k^2 A_k} \propto \frac{1}{\sqrt{m}}\frac{U_j^2 A_j }{\sum_k U_k^2 A_k},
\end{align}\end{linenomath}
which reminds us of the Wiener-Khinchin's theorem that the power is proportional to square of the Fourier amplitude of the mode. In fact, in the case of $m=1$ it holds $A_j=1$ so that the power is proportional to the square of the amplitude $U_j$. In a mono-cyclic signal, i.e. where all $k=j$, the power of the resonance peak will decrease approximately in inverse proportion to square root of the sampling interval $\frac{1}{\sqrt{m}}$. 

\begin{figure}
\begin{center}
\subfigure[]{
\includegraphics[width=0.2\columnwidth]{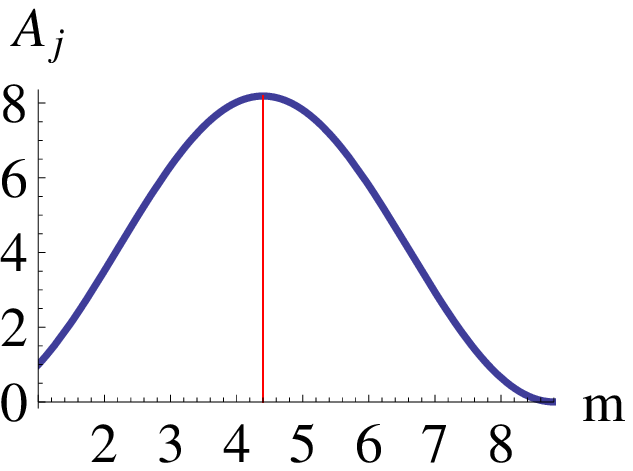}}%scale=0.3
\subfigure[]{
\includegraphics[width=0.2\columnwidth]{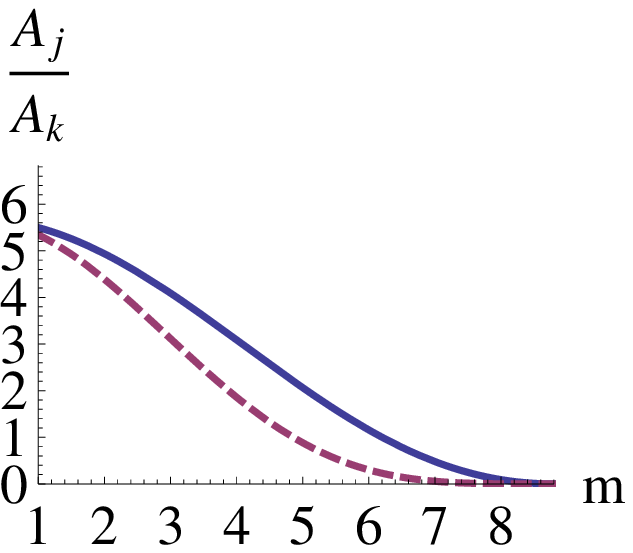}}%\\
\subfigure[]{
\includegraphics[width=0.2\columnwidth]{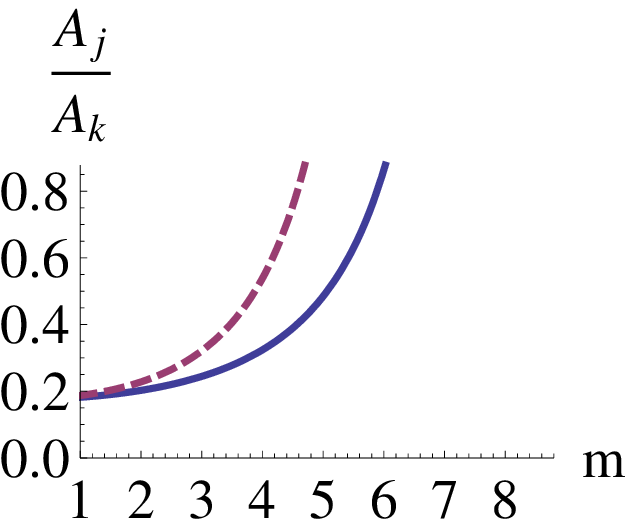}}
\subfigure[]{
\includegraphics[width=0.2\columnwidth]{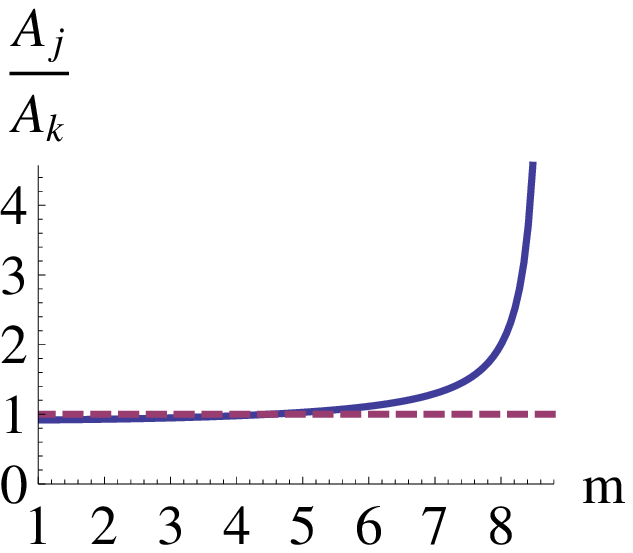}}
\caption{\label{fig:Aratioapp} \small The variation of $A_j$ in $T_j=88$ (a) and of the ratio $\tfrac{A_j}{A_k}$ in (b) $T_j=88,\  T_k=210$, (c) $T_j=210,\ T_k=88$ and (d) $T_j=92,\ T_k=88$ vs. sampling interval $m$. In all cases $\Delta t=1$ and $N=200$. In (a) the vertical bar stands for the Nyquist sampling, $m=\tfrac{T_j}{2\Delta t}$. In (b), (c) and (d) the solid lines stand for the exact evaluation (l.h.s. of $\approx$) and the dashed lines for the approximate evaluation (r.h.s. of $\approx$) for the ratio $\frac{A_j}{A_k}$ in Eq.~\eqref{eq:Aratioapp}.}
\end{center}
\end{figure}

\begin{figure}
\begin{center}
\subfigure[]{
\includegraphics[scale=0.4]{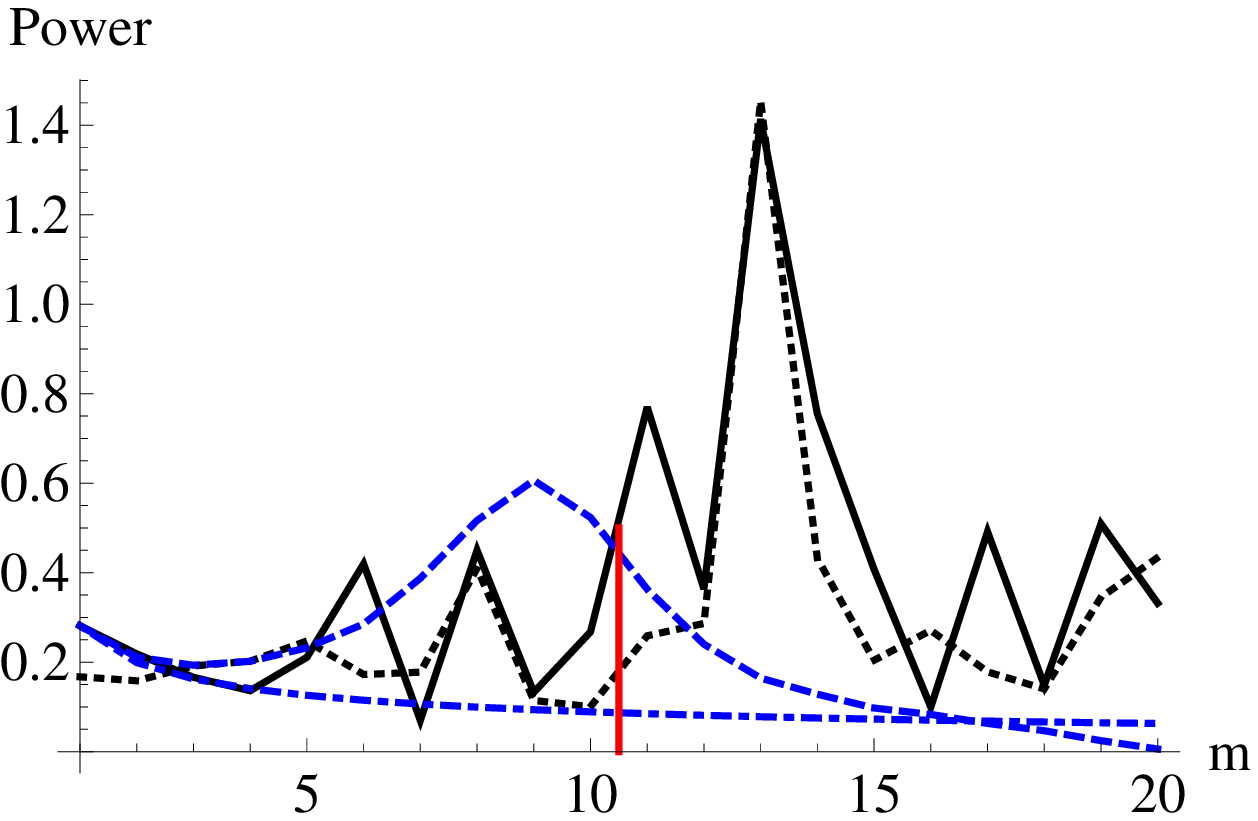}}
\subfigure[]{
\includegraphics[scale=0.4]{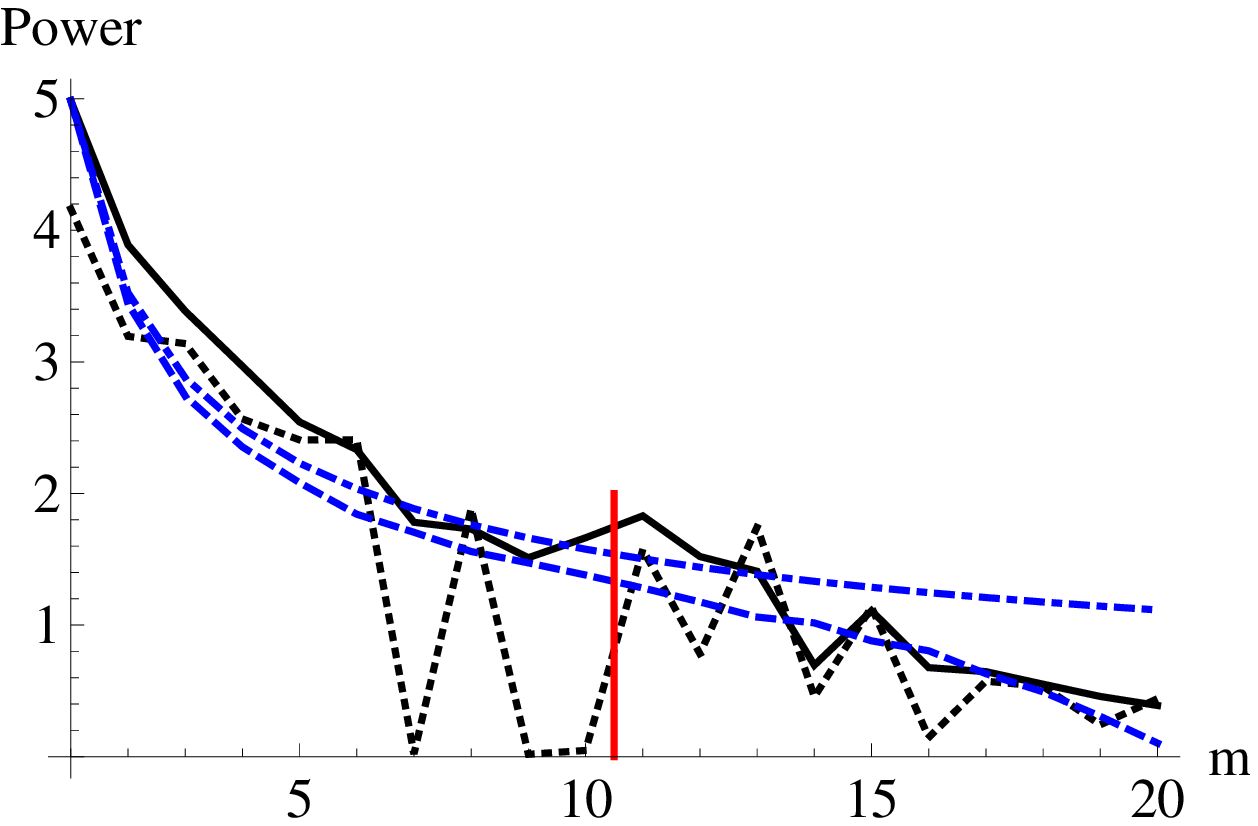}}
\subfigure[]{
\includegraphics[scale=0.4]{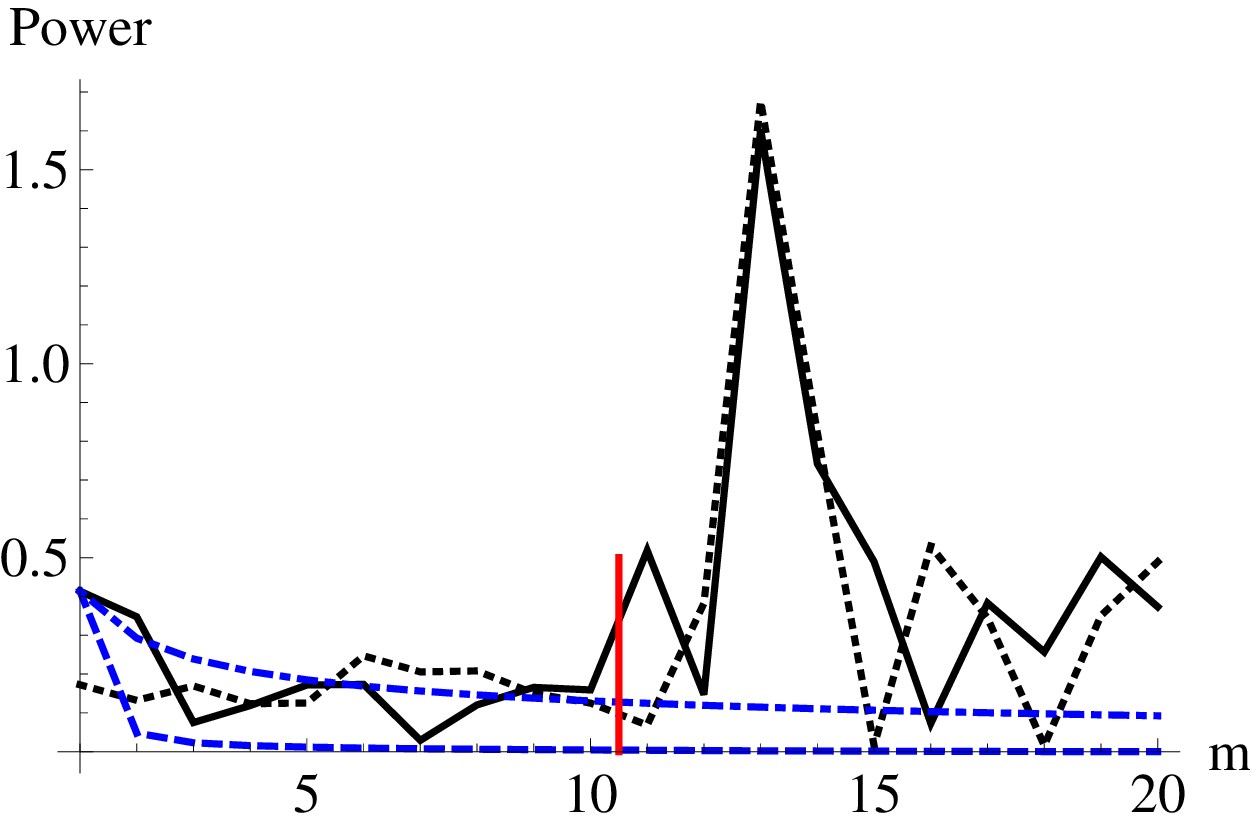}}
\caption{\label{fig:wghcent} \small The variation of the power for the $j=2$-nd mode in a composite signal consisted of three modes: $T_1=95, T_2=210, T_3=330, \phi_1=1.5, \phi_2=1, \phi_3=2.4, \Delta t=10 \text{ and }N=200$ in the cases of (a) $U_1=10, U_2=2, U_3=3$, (b) $U_1=1, U_2=10, U_3=3$, (c) $U_1=1, U_2=2, U_3=10$.. The variations of its amplitudes are different.  The solid line indicates numerical evaluation, the dotted line -- for the approximate evaluation by Eq.~\eqref{eq:psfinal}, the dash-dotted line for the variation proportional to $\frac{1}{\sqrt{m}}$ and the dashed line for the variation according to Eq.~\eqref{eq:normdec}. The red vertical bar stands for the Nyquist sampling $m\Delta t=\frac{T_2}{2}$. The peaks after the Nyquist sampling are nodal peaks.}%(See \S2.2%~\ref{subsec:ridges}\ in Paper II.)}
\end{center}
\end{figure}

However, in the multi-cyclic signal, the sampling (i.e. $m$) dependency will be complicated due to that $A_j$ depends also on sampling interval $m$. Eq.~\eqref{eq:Ajq} for $A_j$ can be rewritten:
\begin{linenomath}\begin{align} \label{eq:Ajmod}
A_j=\frac{1-\exp \left( \dfrac{ 2 \pi  i q_{0j}}{n}\right) }{1-\exp \left( \dfrac{ 2 \pi  i q_{0j}}{mn}\right)} \;
\frac{1-\exp \left( -\dfrac{ 2 \pi  i q_{0j}}{n}\right) }{1-\exp \left( -\dfrac{ 2 \pi  i q_{0j}}{mn}\right)} 
\approx \frac{1-\exp \left( \dfrac{ 2 \pi  i m \Delta t}{T_j}\right) }{1-\exp \left( \dfrac{ 2 \pi  i \Delta t}{T_j}\right)} \;
\frac{1-\exp \left( -\dfrac{ 2 \pi  i m \Delta t}{T_j}\right) }{1-\exp \left( -\dfrac{ 2 \pi  i \Delta t}{T_j}\right)} %\nonumber
=\frac{\sin^2 \left( \dfrac{ \pi  m \Delta t}{T_j}\right)}{\sin^2 \left( \dfrac{ \pi  \Delta t}{T_j}\right)},
\end{align}\end{linenomath}
where Eq.~\eqref{eq:T_j} is used. A relation of $A_j$ vs. $m$ is shown in Fig.~\ref{fig:Aratioapp}(a). $A_j$ reaches maximum at $m=\tfrac{T_j}{2\Delta t}$ (i.e. the Nyquist sampling of the $j$-th mode) and the maximum value is
\begin{linenomath}\begin{align}
A_{j \texttt{max}}=\sin^{-2} \left(\dfrac{ \pi  \Delta t}{T_j}\right). 
\end{align}\end{linenomath}
In denominator of Eq.~\eqref{eq:normdec} the sum $\sum_k U_k^2 A_k$ can be replaced with a representative mode $U_k^2 A_k$. A weighted mean mode or predominant mode can be the representative mode. Then Eq.~\eqref{eq:normdec} is reduced to a simple ratio $\frac{A_j}{A_k}$, which is evaluated with approximation:
\begin{linenomath}\begin{align} \label{eq:Aratioapp}
\frac{A_j}{A_k} \propto \frac{\sin^2 \left(\frac{ \pi m \Delta t}{T_j}\right)}{\sin^2 \left(\frac{\pi m \Delta t}{T_k}\right)} \approx 
\begin{cases} 
\left[\dfrac{T_k}{T_j}\dfrac{1+\cos \left(\frac{ \pi m \Delta t}{T_j}\right)}{2}\right]^2 & \text{ over $m\in [1,\frac{T_j}{\Delta t}]$,\qquad if $T_j<T_k$}\\
\left[\dfrac{T_j}{T_k}\dfrac{1+\cos \left(\frac{ \pi m \Delta t}{T_k}\right)}{2}\right]^{-2} & \text{ over $m\in [1,\frac{T_k}{\Delta t}]$,\qquad if $T_j>T_k$} \\
\qquad\qquad 1& \text{ over $m\in [1,\frac{T_j}{\Delta t}]$,\qquad if $T_j \approx T_k$}
\end{cases} 
\end{align}\end{linenomath}
This expression is a rough estimation that, however, shows that the power of longer cycles than the representative mode will vary exceeding the proportion to $\frac{1}{\sqrt{m}}$ and the power of shorter cycle -- exceeded by the proportion to $\frac{1}{\sqrt{m}}$ (Fig.~\ref{fig:Aratioapp}). This can be understood by noting that a decay of short mode at its Nyquist sampling will let the power of longer modes to increase due to the energy conservation of the signal, which belongs to the normalized autocorrelation Eq.~\eqref{eq:ac1} where the power in Eq.~\eqref{eq:normdec} represents a relative fraction of every mode among the whole energy of the signal. If the case is for a mono-cyclic signal or the predominant mode, the power decreases in proportion to $\frac{1}{\sqrt{m}}$. In Fig.~\ref{fig:wghcent} we can see the different sets of modes. If a short-period mode ($j=1$-st mode in Fig.~\ref{fig:wghcent}) is predominant, then the variation of the power for the longer modes(the $j=2$-st mode in Fig.~\ref{fig:wghcent}) will have an elevation around the Nyquist sampling above the proportionality to $\frac{1}{\sqrt{m}}$. If a long-period mode(the $j=3$-rd mode in Fig.~\ref{fig:wghcent}) is predominant, we see the opposite variation for the shorter modes. And for the predominant cycle itself, its degradation when increasing the sampling interval is proportional to $\frac{1}{\sqrt{m}}$.

\section{The samplogram and D-samplogram}\label{sec:smplgrm}
In this section we show the general configuration of the samplogram. We can also the D-samplogram and the effect of differencing. 

\subsection{The configuration of samplogram}\label{subsec:cnfgsmp}
\begin{figure}
\centering
\scalebox{0.4}{\includegraphics{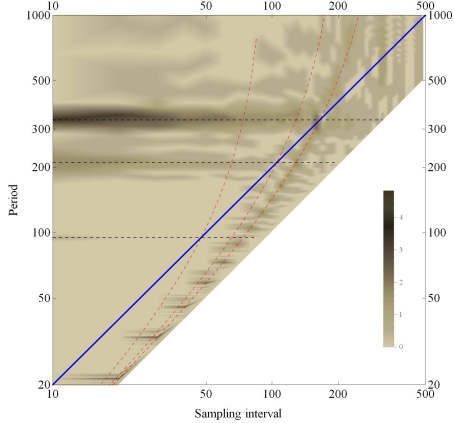}}
\caption{\label{fig:D0} \small The samplogram of a multi-cyclic signal of $U_1=1, U_2=2, U_3=3, T_1=95, T_2=210, T_3=330, \phi_1=1.5, \phi_2=1, \phi_3=2.4, \Delta t=10 \text{ and }N=200$. The solid blue line means the Nyquist sampling where the sampling interval is half of the period. The diagram extends to the Fourier limit where the period equals the sampling interval. The dashed black horizontal lines stand for the true ridges and the dashed red curved lines for aliasing ridges. A sampling peak appears clearly at the Nyquist sampling.}
\end{figure}

\textit{A samplogram} is the diagram that shows how the power changes when the sampling interval increases in the averaging down-sampling. The samplogram has 2 dimensions: \textit{resonance dimension} (period or frequency) and \textit{sampling dimension} (sampling interval). We can consider a third dimension, which is just the power. As upright on the paper, the power can be represented by density or contours. Fig.~\ref{fig:D0} shows a samplogram for a multi-cyclic signal Eq.~\eqref{eq:startsignal} and its sampled time series Eq.~\eqref{eq:oritmsrmlti}, where $U_1=1, U_2=2, U_3=3, T_1=95, T_2=210, T_3=330, \phi_1=1.5, \phi_2=1, \phi_3=2.4, \Delta t=10 \text{ and }N=200$.

In samplogram Fig.~\ref{fig:D0} the Nyquist sampling (solid blue) and the Fourier sampling limit (oblique right boundary of the diagram) are visible. The samplogram is a version of the power spectrum in two dimensions. Also, the power spectrum is the Fourier transform of autocorrelation. In the Fourier transformation of a band-limited signal, the period that is considered is limited down to the sampling interval including the aliasing region. In other words, once a sampling interval is given, cycles of period less than the sampling interval are not considered. This sets the Fourier sampling limit to which the samplogram extends. 
The Nyquist sampling corresponds to that the sampling interval is half of the period of cycle. The sampling theorem says that a cyclic signal can be fully reconstructed only if it is sampled by an interval less than half of the period. As seen at the end of \S~\ref{subsec:apprps}, there appears the sampling peak at the Nyquist sampling. 
The samplogram is bisected by the Nyquist sampling. The region between $m=1$, that means no down-sampling and the ordinate, and the Nyquist sampling is called \textit{a true periodicity region} while the region between the Nyquist sampling and the Fourier sampling limit is called \textit{an aliasing periodicity region}. Ordinary power and Fourier spectra should deal only with the true periodicity region while the samplogram shows both regions.

In Fig.~\ref{fig:D0} two types of ridges are shown: true ridges and aliasing ridges. Both ridges correspond to extensions of the two types of peaks: the true peak and the aliasing peak. When increasing the sampling interval, tracks of peaks form \textit{the ridges}. Both ridges can be located in the frequency domain as follows:
\begin{linenomath}\begin{align}
q&=q_{0j}, \qquad	&\text{for the true ridge}\\
q&=n-q_{0j},\qquad\	&\text{for the aliasing ridge}
\end{align}\end{linenomath}
where $n$ is the length of the down-sampled time series $u_{\bar{m}}$ and $q_{0j}$ is the frequency number for the $j$-th mode. In the period domain the above equations can be rewritten:
\begin{linenomath}\begin{align} \label{eq:ridgeper}
T&=\frac{m n\Delta t}{q_{0j}},\qquad &\text{for the true ridge} \\
T&=\frac{m n\Delta t}{n-q_{0j}},\qquad &\text{for the aliasing ridge}
\end{align}\end{linenomath}
where we recall 
\begin{linenomath}\begin{align}\label{eq:ridgeass}
n&=\texttt{IntegerPart}\left(\frac{N}{m}\right)\approx \frac{N}{m}, \\
q_{0j}&=\texttt{Round}\left( \frac{m n \Delta t}{T_j}\right)\approx \frac{N \Delta t}{T_j}.
\end{align}\end{linenomath}

The same kind of ridges of the different modes do not cross each other: the true ridges of the $j$-th and $l$-th modes as well as their aliasing ridges are parallel. However, the true ridge of a mode and the aliasing ridge of another mode intersect each other. At their crossing a new kind of peak appears, which is called \textit{a nodal peak}. (See Fig.~\ref{fig:Dl}. Don't confuse nodal peaks with discrete peaks in the aliasing ridges, which are related to the sampling discretization.) The sampling peak can be considered a kind of nodal peak where the true and aliasing ridges of the same mode are overlapped. From Eq.~\eqref{eq:psfinal} the power of nodal peak of the $j$-th and $l$-th modes can be expressed
\begin{linenomath}\begin{align}\label{eq:psnod}
PSm(q_{0j},q_{0l})&\approx \frac{\sqrt{n}}{2\sum_k U_k^2 A_k} \left[U_j^2 A_j \exp\left(\frac{2 \pi  i q_{0j}}{n}\right)+U_l^2 A_l \exp\left(-\frac{2 \pi  i q_{0l}}{n}\right)  \right].
\end{align}\end{linenomath}
The nodal peaks are seen in Fig.~\ref{fig:smppkn} as well as Fig.~\ref{fig:wghcent}, which corroborates that the nodal peaks appears on the ridge. From Eqs.~\eqref{eq:ridgeper} and \eqref{eq:ridgeass} the nodal peak of the true ridge of the $j$-th mode and the aliasing ridge of the $k$-th mode happens at the sampling step
\begin{linenomath}\begin{align}\label{eq:nodplace}
m_{jl}\approx \frac{N}{q_{0j}+q_{0l}}.
\end{align}\end{linenomath}
We can see that this location is symmetric for $q_{0j}$ and $q_{0l}$. Thus it is clear that the nodal peak of the true ridge of the $j$-th mode and the aliasing ridge of the $l$-th mode and another nodal peak of the $j$-th aliasing ridge and the $l$-th true ridge happen at the same sampling interval. We can also find the both nodal peaks have the same power. The true ridge and aliasing ridge for the same cycle have the same degradation in terms of the sampling interval. So, the both nodal peaks that appear at the same sampling will have the same power. 

The samplogram can explain the aliasing phenomenon that at lower sampling frequency than the Nyquist frequency an aliasing peak appears in the power and Fourier spectra \citep[e.g.][]{Poluianov2014}. In Fig.~\ref{fig:D0} it is visible that the aliasing ridge of the $j=3$-rd mode invades  into the true periodicity region from the aliasing region. This means that if the sampling rate is lower than the Nyquist rate for the $j=3$-rd mode, then an aliasing peak appears in spectrum while the true peak disappears so that the frequency of the $j=3$-rd mode could be estimated faulty. In fact, the ordinary spectra deal with only the true periodicity region so this aliasing phenomenon was difficult to show. The samplogram shows this intuitively.

\subsection{D$l$-samplogram}\label{subsec:Dlsmplgrm}
\begin{figure}
\begin{center}
\subfigure[]{
\includegraphics[scale=0.4]{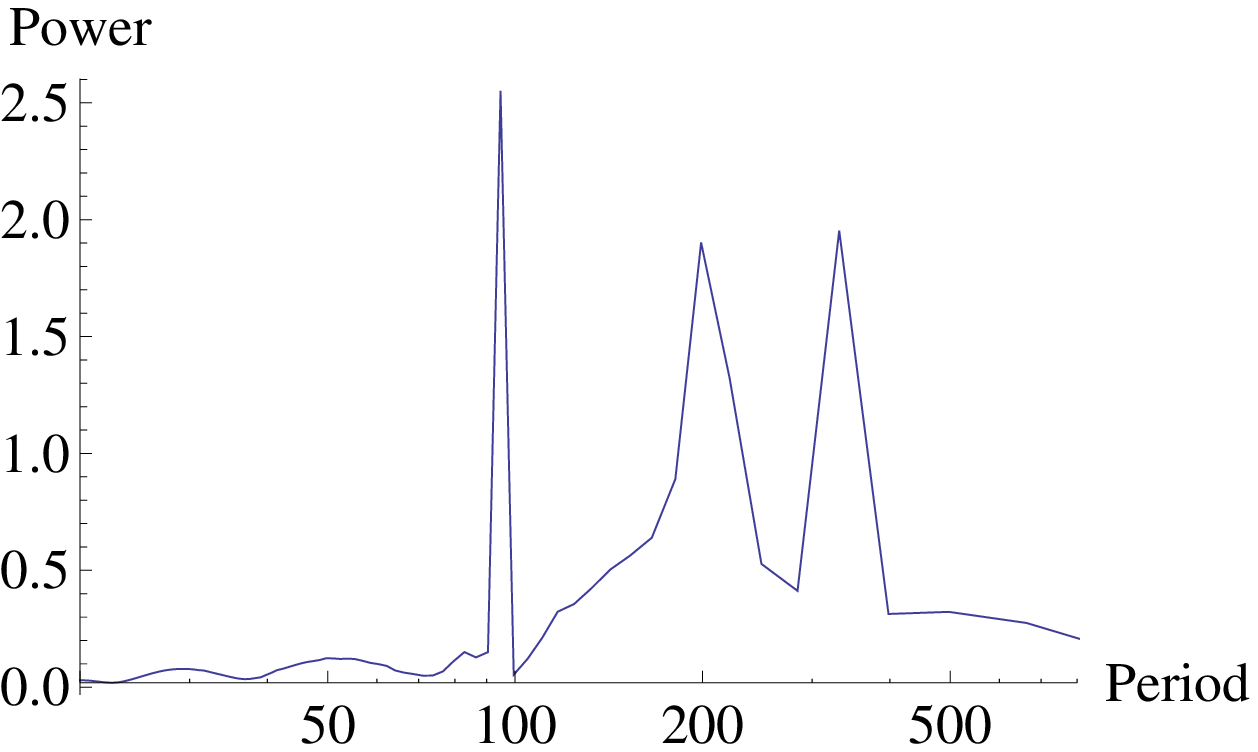}}
\subfigure[]{
\includegraphics[scale=0.4]{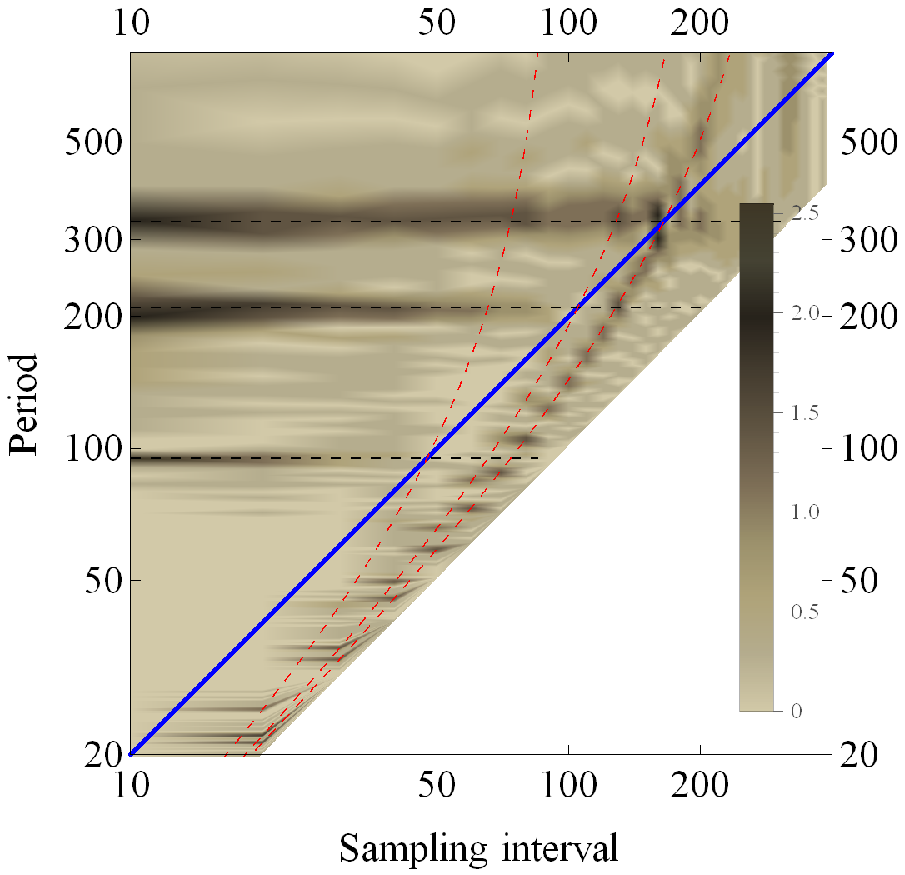}}\\
\subfigure[]{
\includegraphics[scale=0.4]{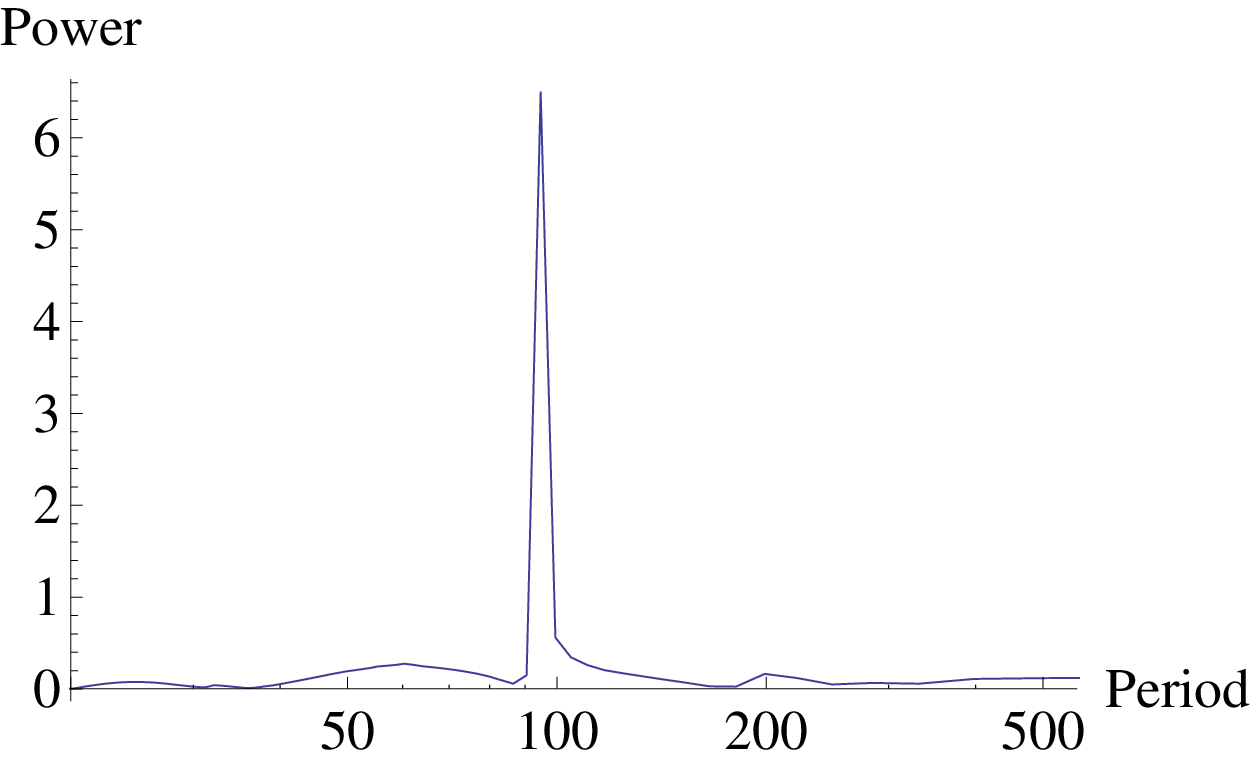}}
\subfigure[]{
\includegraphics[scale=0.4]{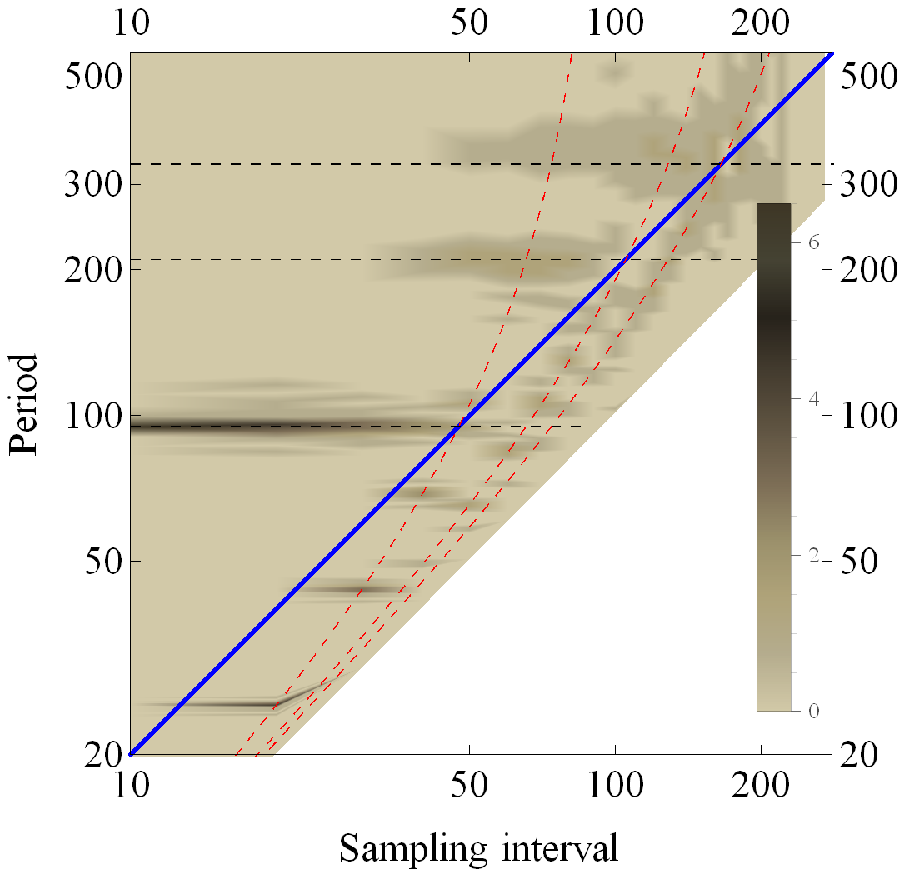}}
\caption{\label{fig:Dl}\small The power spectra and D$l$ (more precisely, $\bar{\text{D}}l$, i.e. the samplograms are evaluated by first averaging and later differencing)-samplograms of the $l$-th order difference series of the multi-cyclic signal in Fig.~\ref{fig:D0}. (a) and (b) show the power spectrum and D$1$ (more precisely, $\bar{\text{D}}1$)-samplogram of the first-order difference series while (c) and (d) -- power spectrum and D$3$ (more precisely, $\bar{\text{D}}3$)-samplogram of the third-order difference series. In power spectra the longer cycles are the more depressed. Dashed lines indicate the ridges. At crossing of ridges, the nodal peaks emerge. }
\end{center}
\end{figure}

In most physical processes, the difference series is considered to clarify the underlying dynamics. The Brownian motion is an typical instance where the displacement but the position itself is the basic dynamic quantity. The difference series consists of differences between adjacent points of the time series:
\begin{linenomath}\begin{align}
\Delta u(r)=u(r+1)-u(r)\qquad \text{for a time series }u(r), 
\end{align}\end{linenomath}
where $r$ is the index of point. If the original time series is a multi-cyclic one $u(r)$ in Eq.~\eqref{eq:oritmsrmlti}, then its difference series can be written 
\begin{linenomath}\begin{align}\label{eq:diftmsr}
\Delta u(r)&= \sum _j  \frac{2 \pi \Delta t}{T_j} U_j \cos \left(\frac{2 \pi r \Delta t}{T_j}+\phi _j+\frac{\pi}{2}\right).
\end{align}\end{linenomath} 
Here we can be sure that the periodicity in time series conserves by the differencing, which is just \textit{a differencing stability of cycle}. And we can find that the amplitude of a mode decreases in inverse proportion to its period. So the differencing can be considered a kind of long-period (low- frequency) filter. If the series is $l$th-order differentiated, a stronger filter will be given:
\begin{linenomath}\begin{align}\label{eq:diftnmsr}
\Delta^l u(r)&= \sum _j  \left(\frac{2 \pi \Delta t}{T_j}\right)^l U_j \cos \left(\frac{2 \pi r \Delta t}{T_j}+\phi _j+\frac{l\pi}{2}\right).
\end{align}\end{linenomath} 
According to the Wiener-Khinchin's theorem, the power of a mode should decrease in inverse proportion to the $2l$-th power of its period in the $l$-order differencing (see Fig.~\ref{fig:Dl}). If the period of a mode is long enough, its power might become neglected at all. Furthermore, according to the regular degradation in Eq.~\eqref{eq:normdec}, the ridges would be degenerated when increasing sampling interval. So, at the Nyquist sampling the power could be more negligible than a resonance peak at $m=1$.

However, Fig.~\ref{fig:Dl} (in particular (d)) shows that the ridges are elevating just before the Nyquist sampling for the differences. Why are the ridges of negligible modes in power spectra elevated when approaching its Nyquist sampling? It can be explained from that the differencing is acting like a long-period (or low-frequency) filter while the Nyquist sampling acting like a short-period (or high-frequency) filter. Consider a mode, e.g. the $j=2$-th mode in Fig.~\ref{fig:Dl}. At the Nyquist sampling of this mode, the shorter $j=1$-st mode has already passed well beyond its Nyquist sampling so the mode would decay at all. On the other hand, through the differencing, the longer $j=3$-rd mode will be depressed much enough. Thus at the Nyquist sampling of the $j=2$-th mode, only the $j=2$-th mode itself will remain, in spite of that its power is negligible in power spectrum of $m=1$ and the 1-st and 3-rd modes are depressed. So, the power of the $j=2$-th mode will be increased up at its Nyquist sampling.
This fact can be explained in another way. As said in \S~\ref{subsec:powvar}, if a short cycle is most significant in signal, the degradation of the longer cycles have an elevation just around the their Nyquist sampling (in comparison with the proportionality to $\frac{1}{\sqrt{m}}$). 
The effect of differencing is so strong that we can neglect the amplitude of modes in original time series and can reorder the power of modes merely by their period in some high-order differences. From this fact, the diagram can show the successive upheavals of ridges in the order of their period around their Nyquist sampling.

\begin{figure}
\begin{center}
\subfigure[]{
\includegraphics[scale=0.4]{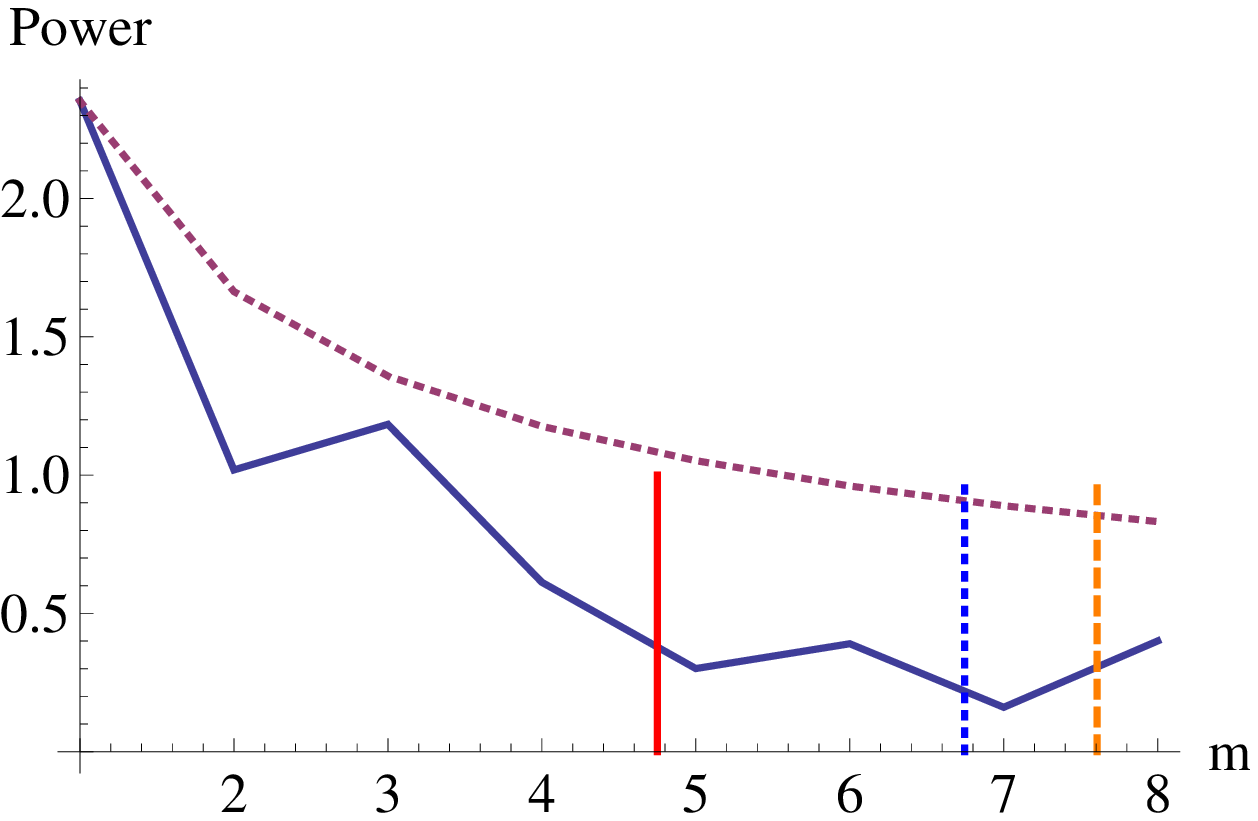}}
\subfigure[]{
\includegraphics[scale=0.4]{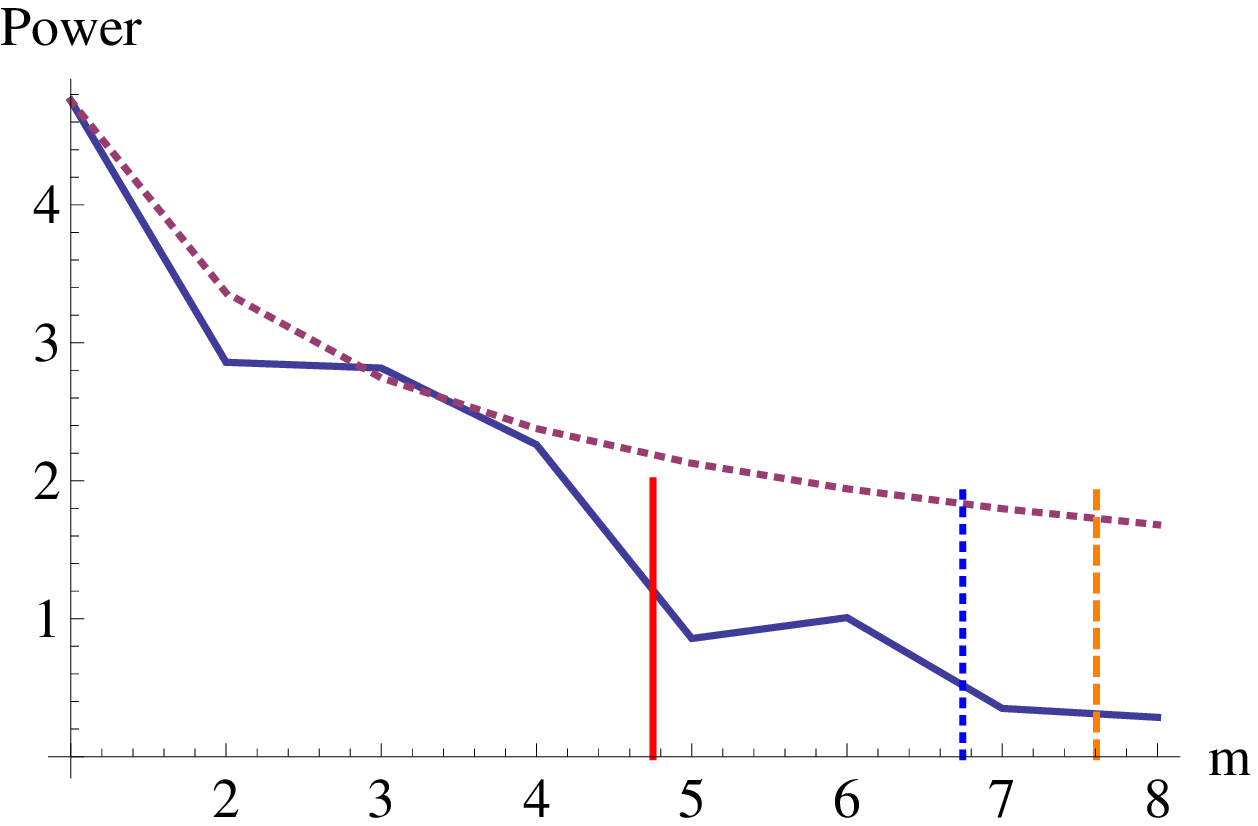}}
\subfigure[]{
\includegraphics[scale=0.4]{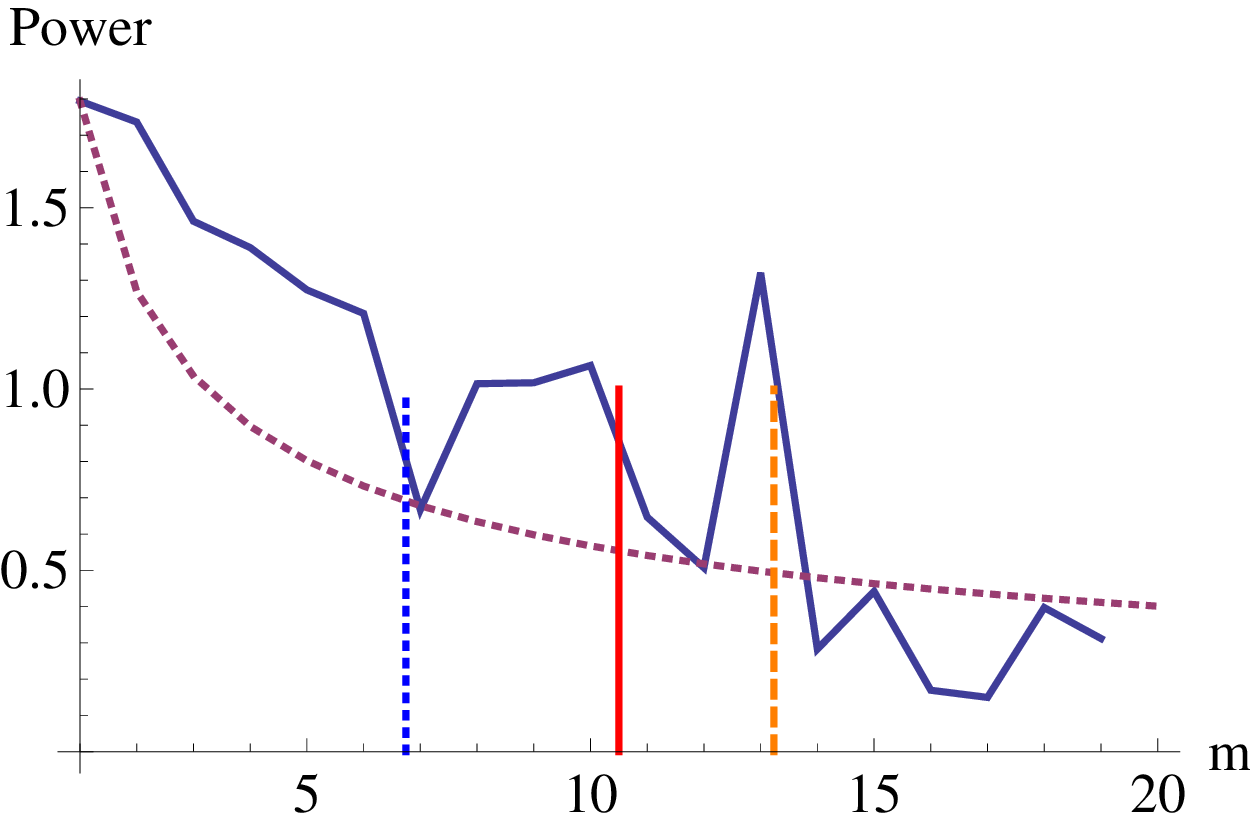}}
\subfigure[]{
\includegraphics[scale=0.4]{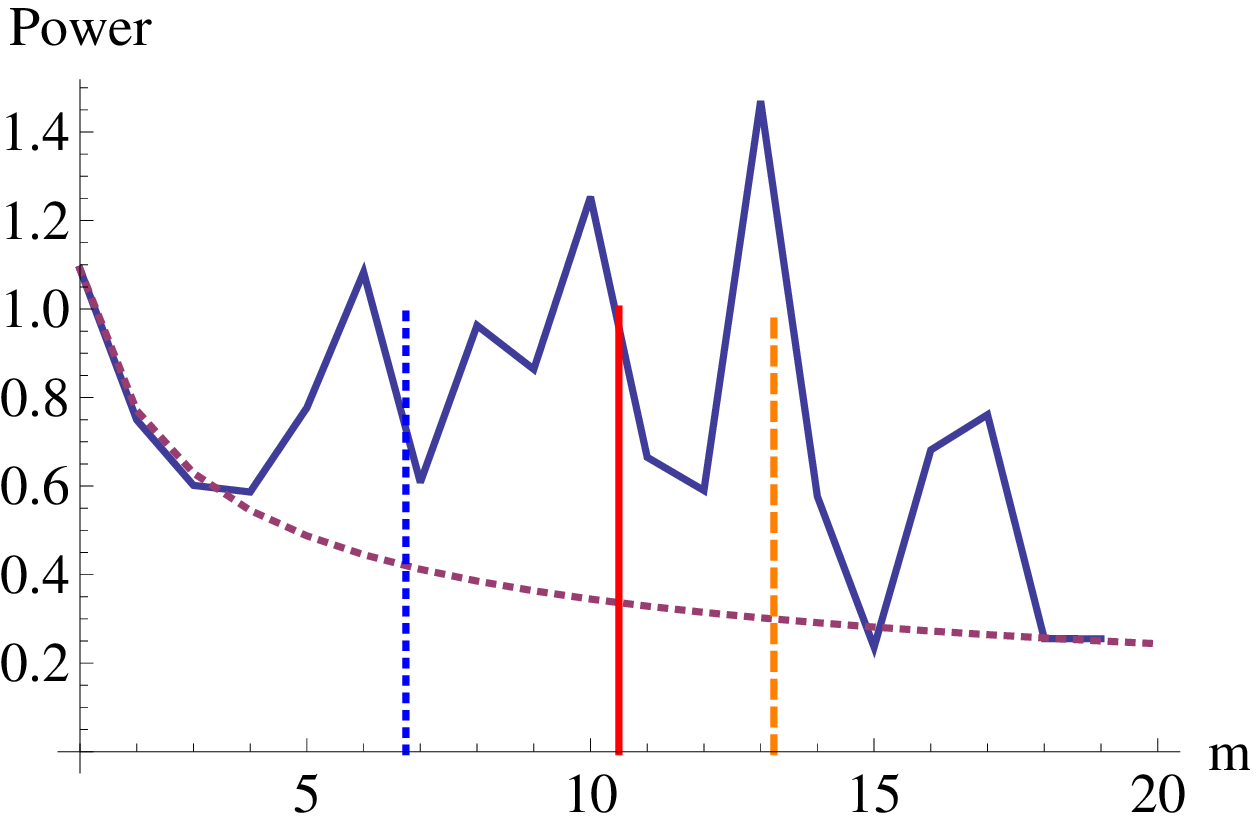}}
\subfigure[]{
\includegraphics[scale=0.4]{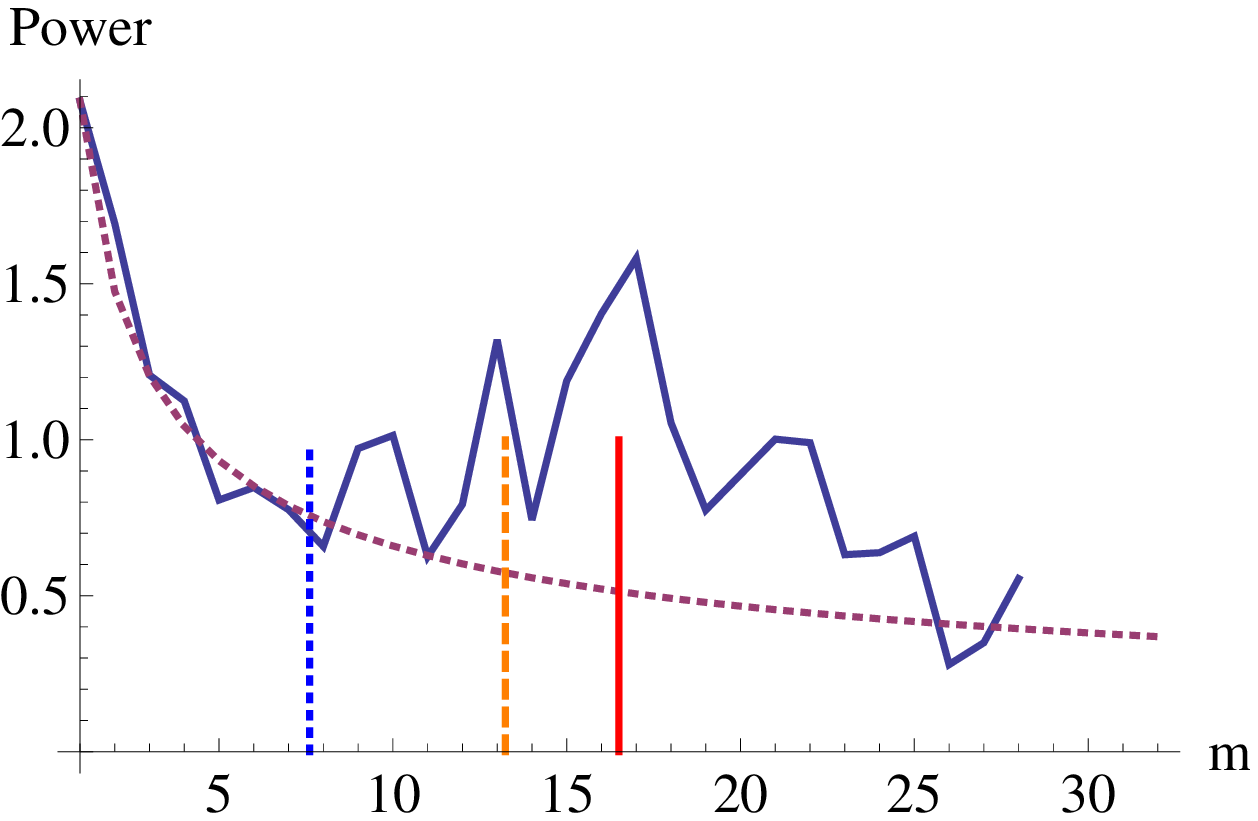}}
\subfigure[]{
\includegraphics[scale=0.4]{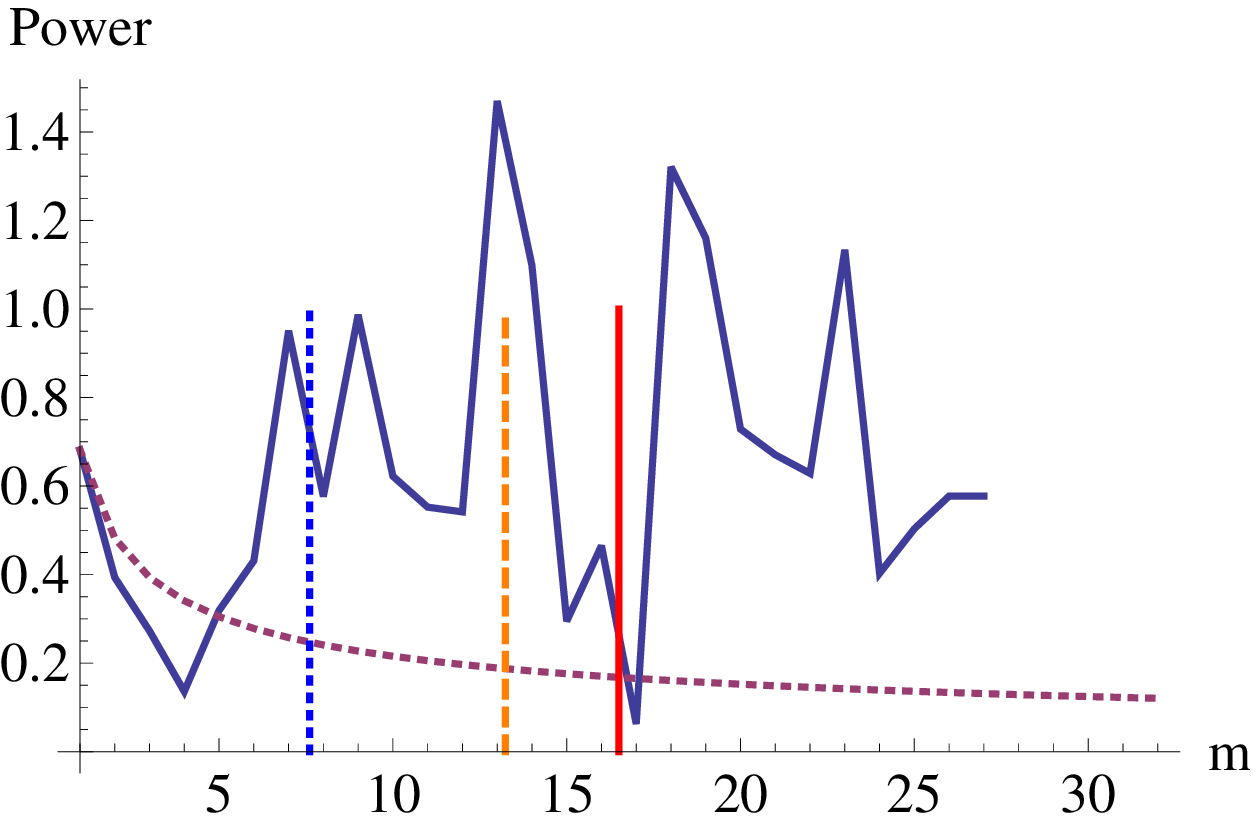}}
\caption{\label{fig:elevNyq}\small The variations of power for modes in Fig.~\ref{fig:D0} in D$1$- and D$2$-samplogram along with sampling interval $m$. (a), (c), (e) correspond to the modes of $j$=1, 2 and 3 in D$1$-samplogram and (b), (d), (f) to the same modes in D$2$-samplogram. The vertical bars mean the nodal and sampling peaks. The peaks look to be shifted due to truncation of series, which is said in Eq.~\eqref{eq:n}. The shortest $j$=1-st mode has shown a ridge degrading below the proportionality to $\frac{1}{\sqrt{m}}$ but longer modes have shown some elevation, in particular, just before the Nyquist sampling. Nodal peaks strengthen such elevations. We can find that the higher the differencing order would be the shorter the period of the predominant cycle should be so that the more cycles should have the elevation around the Nyquist sampling.}
\end{center}
\end{figure}

The nodal peaks swell this elevation more around the Nyquist sampling. In fact, nodal peaks are distributed much more around the Nyquist sampling. Thus, the nodal peaks can exceed the sampling peaks (Fig.~\ref{fig:elevNyq}). The nodal peaks distributed around the Nyquist sampling look like the Laue's X-ray diffraction pattern that is a powerful tool to analyze the structures of the material so we could expect an information of signal from those nodal peaks.

The samplogram accents the behavior of the power at the Nyquist sampling. This behavior looks rather more obvious in the differences. We'd like to call the samplogram of $l$-th order difference series \textit{a D$l$-samplogram}. According to experiences, the D$2$- or D$3$- samplogram will make the  ridges elevating successively in the order of their period around the Nyquist sampling.

The D$l$-samplogram needs both differencing and averaging. Normal procedure consists of first differencing and later averaging. We can take the reverse order of operations: first averaging and later differencing. The reverse procedure shows similar properties as the normal one. We denote this kind of samplogram as $\bar{\text{D}}l$-samplogram. Fig.~\ref{fig:Dl} shows the $\bar{\text{D}}1$- and $\bar{\text{D}}3$-samplogram depicted by the reverse order of operations. The difference between D$l$- and $\bar{\text{D}}l$-samplograms appears for a stochastic signal while they appear to be the same for a deterministic signal.

\section{Enhancing visuality and extending to the Lomb-Scargle periodogram}\label{sec:enhancVis}
We can enhance the visuality of the samplogram such as the resolution, contrast and so on by some sophisticated tricks. We can also extend the samplogram analysis to the well-known Lomb-Scargle periodogram, which will give us more advantages.

\subsection{The post-zero padding and period resolution}\label{subsec:zeropad}
The resolution and contrast in the samplogram can be improved by using some sophisticated methods.

The length of the down-sampled time series decreases inversely proportional to the sampling step $m$ (Eq.~\ref{eq:n}): $n=\texttt{IntegerPart}(N/m)$, which leads to lowering the resolution of period, in particular, at low frequencies. In deed, using the expression for period in DFT (Eq.~\ref{eq:DFTper}), the resolution of period can be defined 
\begin{linenomath}\begin{align}\label{eq:resT}
\Delta T(q)=T(q+1)-T(q)=\frac{N \Delta t}{q+1}-\frac{N \Delta t}{q}=\frac{N \Delta t}{q(q+1)}\approx \frac{N \Delta t}{q^2},
\end{align}\end{linenomath}
where $q$ is the frequency index. If the length $N$ is changed into $N'$, the frequency number $q$ for the given period $T$ is changed:
\begin{linenomath}\begin{align}
q'=\frac{N'\Delta t}{T}=q\frac{N'}{N},
\end{align}\end{linenomath}
from which
\begin{linenomath}\begin{align}
\Delta T'(q)\approx \frac{N' \Delta t}{q'^2}=\frac{N}{N'}\frac{N \Delta t}{q^2}=\frac{N}{N'}\Delta T(q)
\end{align}\end{linenomath}
So the resolution of period is proportional to the length of the time series. To avoid lowering the resolution in the down-sampling, the zero padding can be applied at the end of the autocorrelation series $ACm(g)$ of every down-sampled time series, which is called the post-zero padding. Padding any number of zeros at the end of a signal has no effect on its DFT \citep{Orfanidis2010}, while the length of the series increases and the resolution is improved.
 
Here we consider a question in detail: may the resonance peaks be shifted in spectrum by the zero padding? Evaluate the location of the resonance peak in the frequency and period domain. Recall the principal terms of the power of a sinusoidal signal as in Eq.~\eqref{eq:psfinal}. If the post-zero padding is applied to the series of the autocorrelation $ACm(g)$ so the length of series, $n$, grows to $n'$, then the power is reevaluated as
\begin{linenomath}\begin{align}
PSm'(q)&=\frac{1}{\sqrt{n'}}\sum _{g=1}^{n'} ACm(g) \exp \left( \frac{2 \pi  i (g-1) q}{n'}\right) =\frac{1}{\sqrt{n'}}\sum _{g=1}^{n} ACm(g) \exp \left( \frac{2 \pi  i (g-1) q}{n'}\right) \nonumber \\
&\approx \frac{1}{2\sqrt{n'}\sum_k U_k^2 A_k}\sum_j U_j^2 A_j %\times \nonumber\\&\times
\Bigg[\exp\left(\frac{2 \pi  i q_{0j}}{n}\right) \frac{1-\exp\left(2 \pi  i\left(\displaystyle\frac{q}{n'}+\frac{q_{0j}}{n}\right)n\right)}{1-\exp\left(2 \pi  i\left(\displaystyle \frac{q}{n'}+\frac{q_{0j}}{n}\right)\right)} \nonumber \\
&+\exp\left(-\frac{2 \pi  i q_{0j}}{n}\right) \frac{1-\exp\left(2 \pi  i\left(\displaystyle\frac{q}{n'}-\frac{q_{0j}}{n}\right)n\right)}{1-\exp\left(2 \pi  i\left(\displaystyle \frac{q}{n'}-\frac{q_{0j}}{n}\right)\right)} \Bigg].
\end{align}\end{linenomath}
Then the resonance peak of $m=1$ locates at  
\begin{linenomath}\begin{align}
\frac{q}{n'}=\frac{q_{0j}}{n}\qquad \text{or} \qquad q=q_{0j}\frac{n'}{n},
\end{align}\end{linenomath}
which can be rewritten in the period domain  
\begin{linenomath}\begin{align}
T'_j=\frac{mn'\Delta t}{q}=\frac{mn\Delta t}{q_{0j}}=T_j, \qquad \text{and} \qquad%\nonumber \\
T'_{aj}=\frac{mn'\Delta t}{n'-q}=\frac{mn\Delta t}{n-q_{0j}}=T_{aj},
\end{align}\end{linenomath}
where $T_{j}, T_{aj}$ are the true and aliasing periods of the $j$-th mode. Thus we can conclude that the resonance peaks or the ridges should not be shifted by the post-zero padding, while the resolution of period is improved. However, lengthening the time series will accompany lowering Fourier amplitude or power in inverse proportion to the length. It will be considered in next subsection.

The zero-padding implies us an important thing: a non-stationary signal can make a peak in the spectrum. For example, if only one period or short segment of cycle is existing for a much longer time, this segment of the cycle can make a peak in spectrum, which means spurious cycle because the peak in spectrum means the stationary cycle while the segment means non-stationary. For example, if in some period of solar activity there are only two grand minima, then the separation between them pretends to be a stationary cycle via Fourier analysis, because this approach decompose any signal into the series of stationary sinusoidal signals. 

\subsection{The ridge lifting and peak barring}\label{subsec:lifting}
As seen in Eq.~\eqref{eq:normdec}, the power of a cycle decreases inversely proportional to square root of the sampling step $m$, which makes it difficult to observe the behavior at the Nyquist sampling where the long cycles are much lowered. Besides that, the post-zero padding also lowers the power inversely proportional to square root of extended length $n'$. In order to visualize the power at the Nyquist sampling more clearly, it would be needed to lift the ridge for the great sampling interval. To do this, for example, a lifting proportional to $m$ could be applied to the power of every down-sampled series.

\begin{figure}
\begin{center}
\subfigure[]{
\includegraphics[scale=0.4]{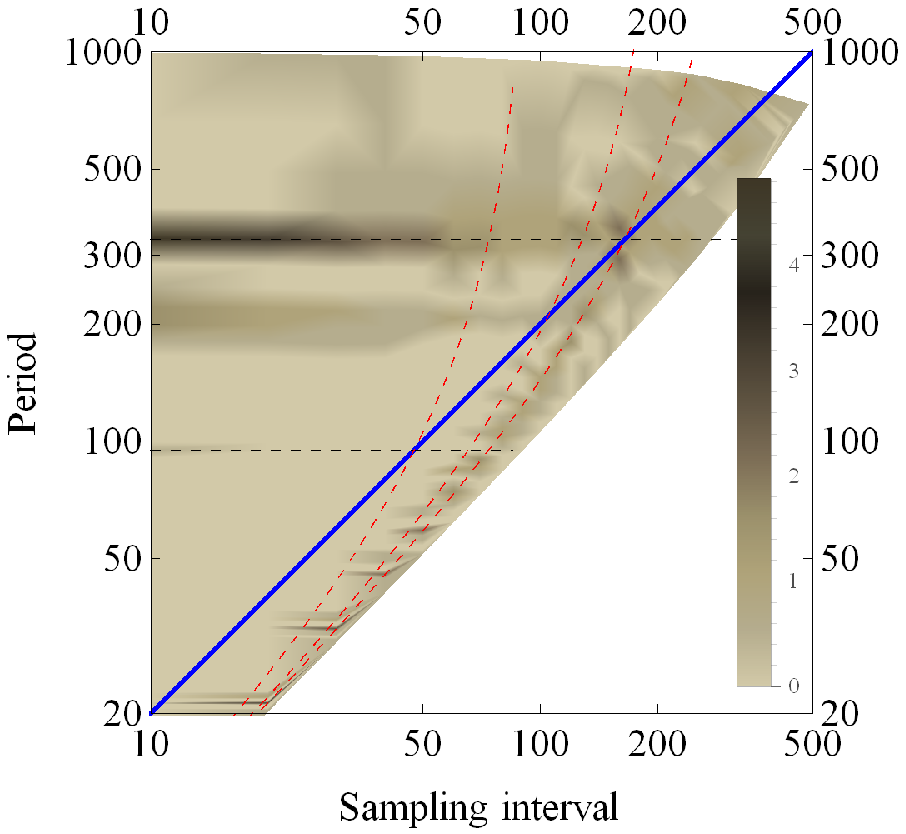}}
\subfigure[]{
\includegraphics[scale=0.4]{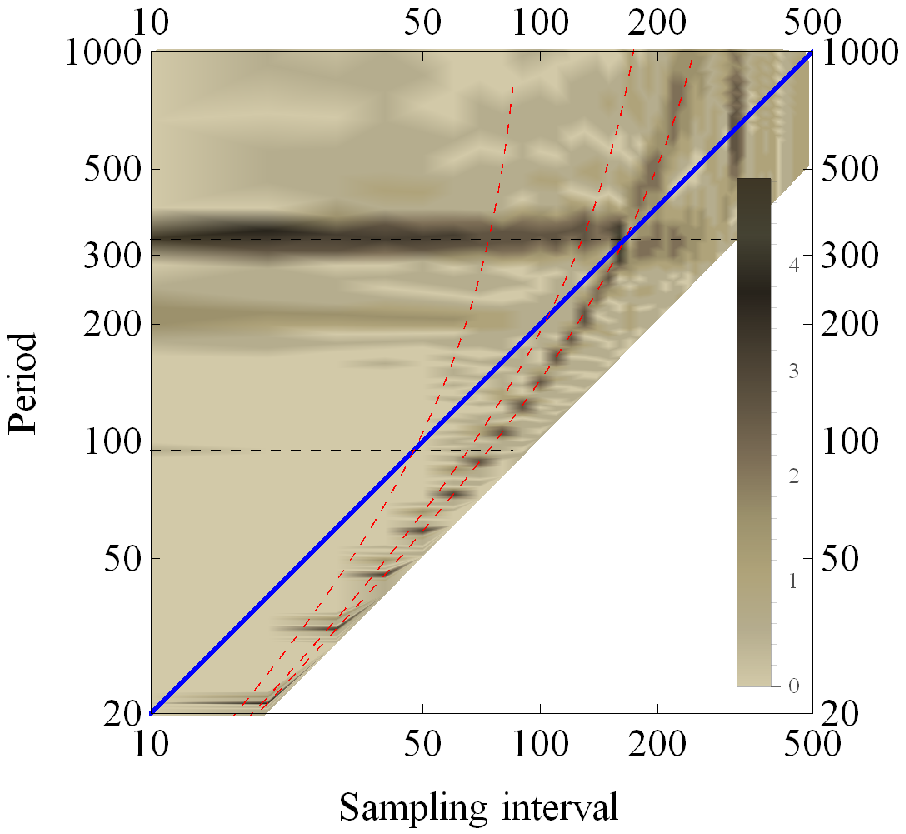}}
\subfigure[]{
\includegraphics[scale=0.4]{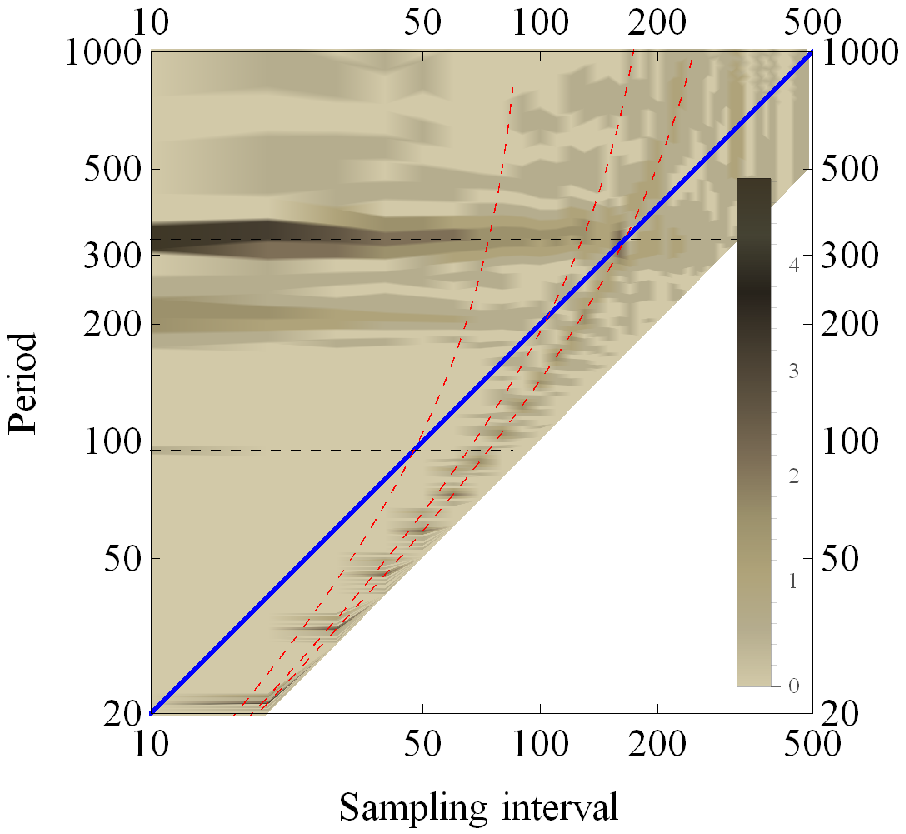}}
\caption{\label{fig:enhancing}\small The raw samplogram (a), visuality-enhanced samplogram (b) with an extending factor $\sqrt{m}$ and a lifting factor $\sqrt[4]{m}$ and the peak-barred samplogram(c) with an extending factor $\sqrt{m}$ and a unit lifting factor.}
\end{center}
\end{figure}

Figure~\ref{fig:enhancing} shows the raw samplogram and visuality-enhanced samplogram with a extending (by zero padding) factor $\frac{n'}{n}=\sqrt{m}$ and a lifting factor $\sqrt[4]{m}$. Manipulation begins with that the series of autocorrelation for every down-sampled series of length $n=\frac{N}{m}$ is so zero-padded that its length becomes $n'=\frac{N}{\sqrt{m}}$, that is, $n'-n$ zeros are padded at the end of time series of autocorrelation. The length of the time series is extended by a factor $\sqrt{m}$ that may be user-defined, where $N$ is the length of the initial time series before the down-sampling and $m$ is averaging step. Lowering power of every down-sampled series due to extending is compensated through multiplication by a factor $\sqrt{\sqrt{m}}=\sqrt[4]{m}$ to the down-sampled time series. Then, the power is multiplied additionally by the lifting factor $\sqrt[4]{m}$ that may also be user-defined. In Fig.~\ref{fig:enhancing}, it is visible that the contrast of ridges against background and the resolution of period for long cycle are enhanced, in particular, at the Nyquist sampling. 

As mentioned earlier, in DFT the peak appears at \texttt{Round}$(q_{0j})$ but not at the decimal frequency number $q_{0j}$, where \texttt{Round}() means the rounding value. The period determined by Eq.~\eqref{eq:ridgeper} may vary when the sampling interval varies. Thus, if we evaluate the period of cycle by a point-like peak, we might find a biased period by rounding, especially in a great sampling interval. 
In fact, if we found the peak at \texttt{Round}$(q_{0j})$, the real frequency number could be in a range of $[\texttt{Round}(q_{0j})-0.5, \texttt{Round}(q_{0j})+0.5)$. So we can depict the samplogram not with point-like peak but with barred peak where the range of $[\texttt{Round}(q_{0j})-0.5, \texttt{Round}(q_{0j})+0.5)$ has all the same power as the peak at frequency \texttt{Round}$(q_{0j})$ (Fig.~\ref{fig:enhancing}c). This will help to find the exact frequency for a cycle, especially for a low-frequency or long-period cycle that has the low resolution of period.

\subsection{Extending to the Lomb-Scargle periodogram}\label{subsec:LSper}
The Lomb-Scargle (LS) periodogram is somewhat different from the aforementioned AC periodogram (Eq.~\ref{eq:ACPer}). This can be applied to the unevenly sampled time series. Suppose that there are $N$ data points $u(r)=u(t_r), r=1,2,\dots,N$. Then the LS periodogram of the time series can be defined by \citep{Press2007}
\begin{linenomath}\begin{align}\label{eq:LSP}
P_N(\omega)\equiv\frac{1}{2\texttt{Var}(u)}\left\lbrace \frac{\left[\Sigma_r(u(r)-\bar{u})\cos \omega (t_r-\tau)\right]^2}{\Sigma_r\cos^2 \omega (t_r-\tau)}+\frac{\left[\Sigma_r(u(r)-\bar{u})\sin \omega (t_r-\tau)\right]^2}{\Sigma_r\sin^2 \omega (t_r-\tau)}  \right\rbrace,
\end{align}\end{linenomath}
where $\omega$ is the frequency while $\bar{u}$ and $\texttt{Var}(u)$ are the mean value and variance of $u(r)$, respectively (Eq.~\ref{eq:xbar} and Eq.~\ref{eq:varx1}). And $\tau$ is defined by
\begin{linenomath}\begin{align}\label{eq:tau}
\tan(2\omega \tau)=\frac{\Sigma_r \sin 2\omega t_r}{\Sigma_r \cos 2\omega t_r},
\end{align}\end{linenomath}
which implies the mean value of the time steps $t_r$ \citep{Scargle1989}. The LS periodogram was induced through the least-square method, so this has been often called the least-square (LS) periodogram. 

\begin{figure}
\centering
\subfigure[]{
\includegraphics[width=0.45\columnwidth]{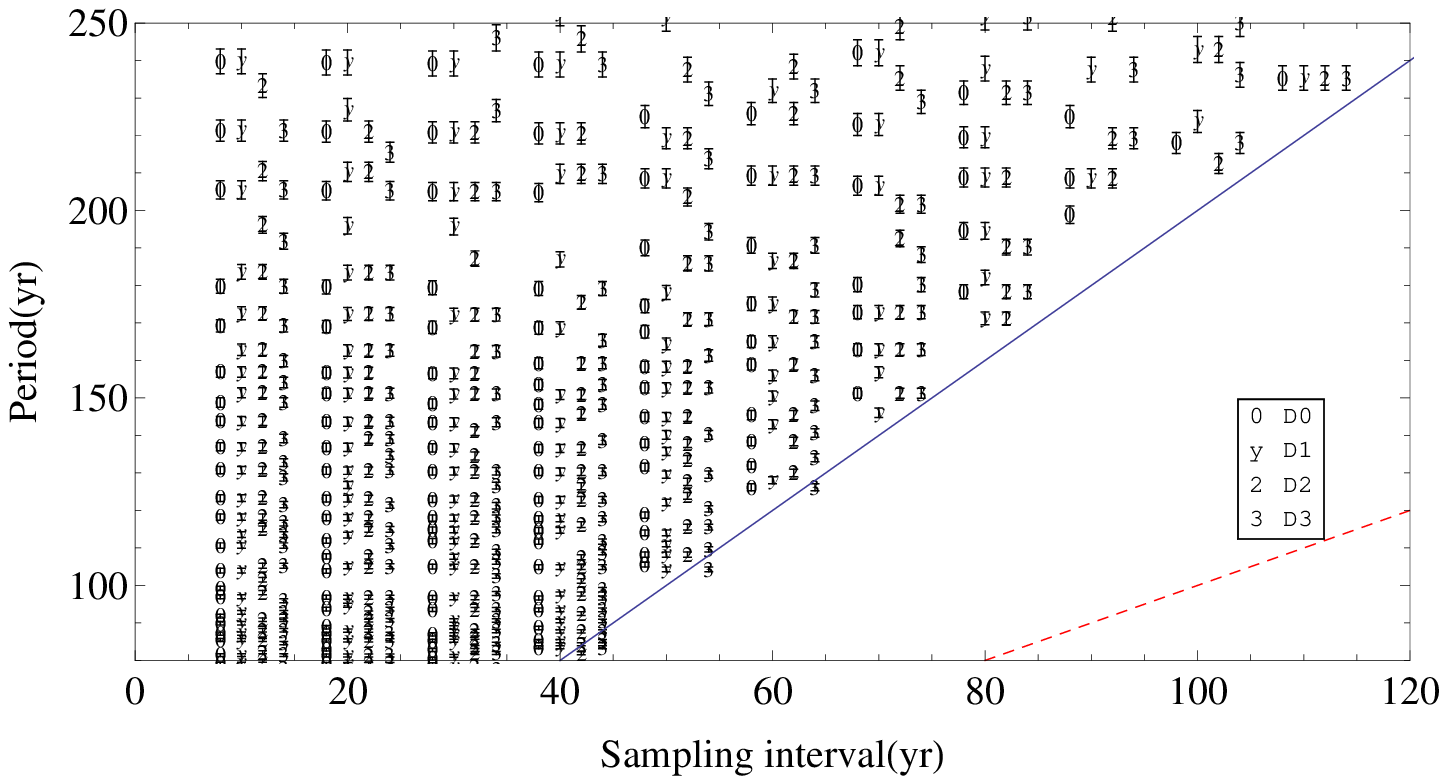}}
\subfigure[]{
\includegraphics[width=0.45\columnwidth]{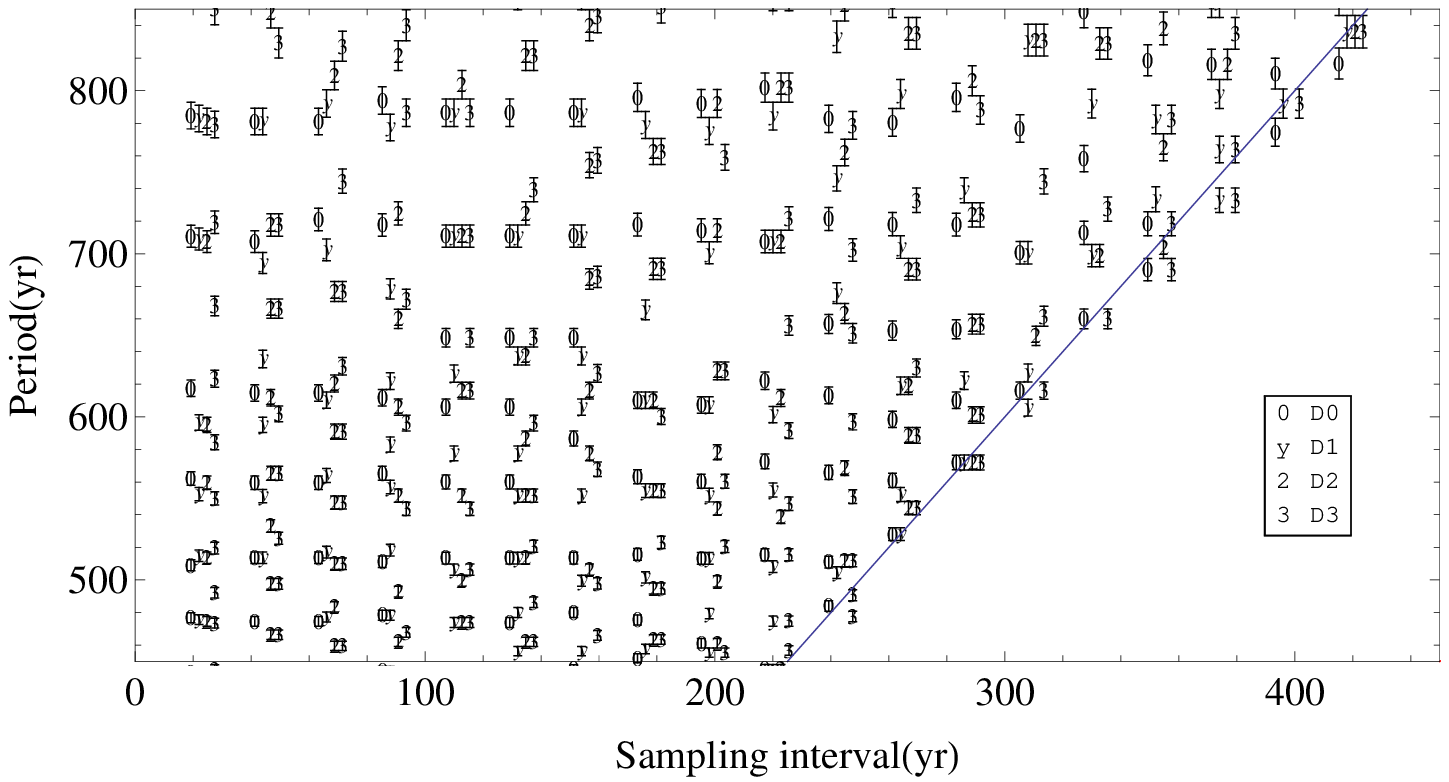}}
 \caption{\label{fig:DstbR} The peaks in the AC samplogram of RSSN within the period of 80 to 250 yr (a) and the peaks in LS samplogram for RTSI within the period of 450 to 850 yr (b) including D-samplograms upto the 3rd order. The marks ``0'',  ``y'', ``2'' and ``3'' stand for the peak points in spectra of the 0-th to 3-rd order differences, respectively. The error bars mean the period resolution related to the rounding of frequency index (cf. the peak barring in \S~\ref{subsec:lifting}). The error bars are displaced horizontally from the original peak points to discriminate the peaks of the different orders. The blue diagonal implies the Nyquist sampling. The region beyond the Nyquist sampling is neglected. In (b) the oversampling factor for LS periodogram is 4.}
\end{figure}

Eq.~\eqref{eq:LSP} appears to give the power at an arbitrary frequency $\omega$ so to be able to provide an high resolution. However, as \citep{Scargle1989} indicated, the number of independent frequencies are limited to the length of time series. So we can not expect higher resolution in LS periodogram in comparison with the classical AC periodogram. However, LS periodogram has a peculiarity that this can be applied to the unevenly sampled dataset, so if we combine the samplogram and LS periodogram, that is if we depict the samplogram in terms of the peaks in LS periodogram, we can expand the samplogram to any unevenly sampled dataset. Figure~\ref{fig:DstbR} shows the peaks in the samplogram of the AC periodogram and LS periodogram for the reconstructed sunspot number (RSSN)\footnote{\label{footnt:SSN}The dataset is available through the MPS sun-climate web-page at \url{https://www2.mps.mpg.de/projects/sun-climate/data/SN_composite.txt}. See also \citet{Wu2018}.} and the reconstruction of total solar irradiance (RTSI)\footnote{\label{footnt:TSI}The dataset is available through the NOAA web-page at \url{https://www1.ncdc.noaa.gov/pub/data/paleo/climate_forcing/solar_variability/steinhilber2012.txt}. See also \citet{Steinhilber2012}.}. 

\section{The practical issues}\label{sec:practice}
So far we have seen only the deterministic periodic signal. The random signal is discussed in this section. And the effectivity of sampling and differencing stability analysis is discussed as well.

\subsection{The features of the sinusoidal signal in the samplogram}\label{subsec:cridetcycle}
According to the idea of Fourier analysis, an arbitrary signal can be linearly decomposed into the series of sinusoidal signals. So it is first important to analyze the sinusoidal signal that we saw in \S~\ref{sec:ps}. The feature of a stationary cycle in the sinusoidal signal in the traditional spectrum is almost unique: 
\begin{itemize}
\item The resonance peak

\textit{The sinusoidal stationary cycle makes a resonance peak in the Fourier or power spectra.} The resonance peak is a peak that appears in the Fourier or power spectra. The resonance peaks also appear in ordinate of the samplogram, i.e. at averaging step $m=1$. As we know, however, many noisy and random peaks pollute the spectra for a random signal and it is difficult to distinguish them.   

\end{itemize}

The samplogram can show the additional features of the sinusoidal stationary cycle, which we have seen before:

\begin{itemize}
\item The ridge

\textit{The sinusoidal stationary cycle makes a ridge in samplogram.} The ridge is an extension of resonance peaks along the sampling interval increasing. The ridge expresses the sampling stability of cycle. 

The sinusoidal signal has a regular degradation of the ridge in the averaging down-sampling. \textit{When increasing the sampling interval or sampling step $m$, the degradation of the longer cycle than the weighted mean or predominant cycle has an elevation in comparison with the proportionality to $\frac{1}{\sqrt{m}}$, while for the shorter cycle a sinking appears. The predominant mode itself degrades in proportion to $\frac{1}{\sqrt{m}}$.}  For a random signal, the randomness is added to this normal degradation, which threatens the existence of peak and ridge.

The elevation of the ridge for the longer cycle before the Nyquist sampling is very special. Peaks that are negligible in the power spectrum become significant before the Nyquist sampling. This property appears especially in the D$l$-samplogram, where the predominant or weighted mean cycle is shifted to higher frequency by differencing and successive elevation of ridges before the Nyquist sampling appears. We can add a virtual ``predominant and very short" cycle to the signal. Then the elevation of ridges for almost all of the cycles are observable including the probable shortest cycle. 

\item The sampling peak

\textit{The sinusoidal stationary cycle makes the sampling peak at its Nyquist sampling.} This peak is formed by superposition of the true and aliasing ridges at the Nyquist sampling. However, in DFT, the appearance of sampling peak is conditional: in the place of this peak a valley might appear. 

Incidentally, the sampling peak has another characteristic which is concerned with the complex argument of power. \textit{At the sampling peak, the power should be a negative real and must have the argument of $\pm\pi$.} In DFT, however, the argument could vary because the frequency number is given by an integer but a decimal number. The argument depends also on the parity of length of time series. 

\item The nodal peak

\textit{In a multi-cyclic signal, different sinusoidal stationary cycles make the nodal peak in the samplogram.} The nodal peak is formed by superposition of a true ridge for one cycle and an aliasing ridge for another cycle. The sampling peak is a kind of nodal peak, which is formed by the true and aliasing ridges for the same cycle. \textit{The two nodal peaks for two cycles appear at the same sampling interval in both true and aliasing periodicity regions and, furthermore, they have the same power.} Although the ridge for cycle may be negligible, the nodal peak on the ridge can appear significant. Like the sampling peak, the appearance of the nodal peak might be conditional.

\end{itemize}

Here the sampling and nodal peaks are conditional, so the ridge can be regarded as a necessary property of the cycle. And the resonance peak is engulfed in the adjacent peak or background, it woulbe better for the existence of cycle to be determined by ridge in samplogram rather than resonance peak in ordinary spectrum. In fact, a negligible cycle in sampling of $m=1$ can become significant at its Nyquist sampling.

The above features have been obtained for a sinusoidal stationary cycle. However, in practice we meet hardly with such a signal. In fact, the solar activity is considered to be a non-stationary and stochastic process. If a random signal is mixed with the deterministic cycles, then what will happen to the above features? Will they be hardly detectable? We consider the random signal and effectivity of the samplogram to detect a weak cycle in it.

\subsection{An example of application to random signal}
\label{subsec:prac}
\begin{figure}
\begin{center}
\subfigure[]{
\includegraphics[scale=0.4]{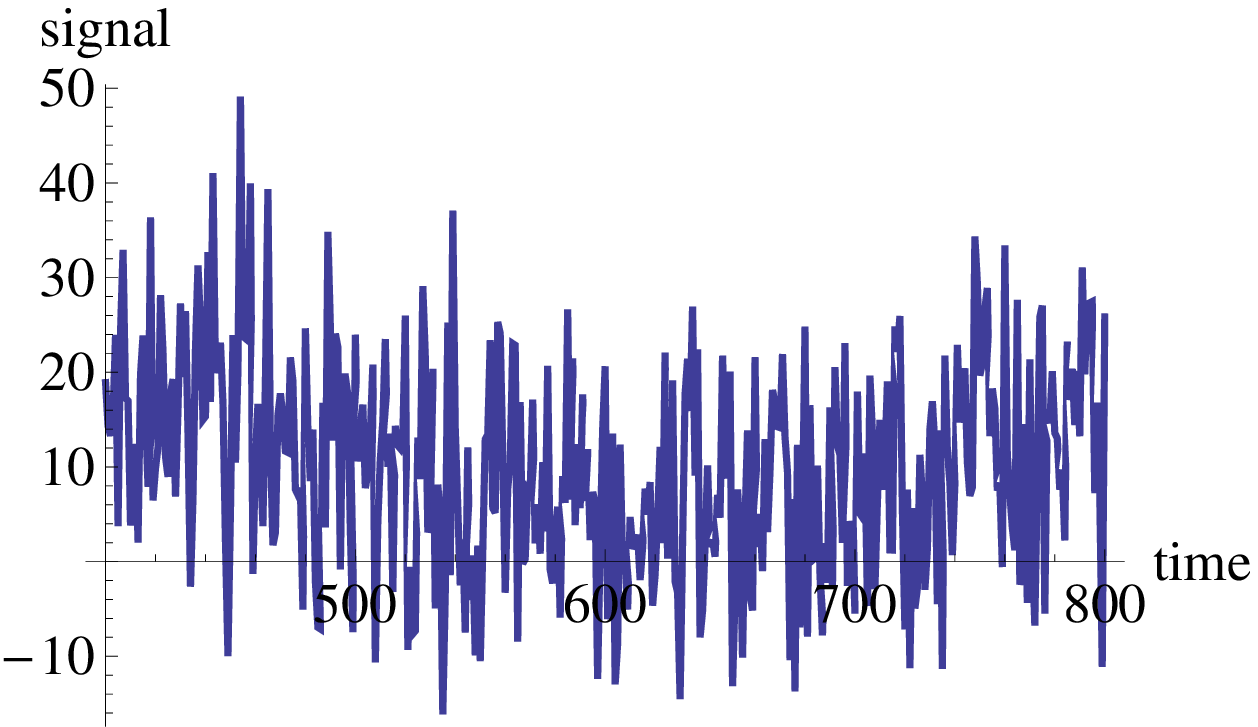}}
\subfigure[]{
\includegraphics[scale=0.4]{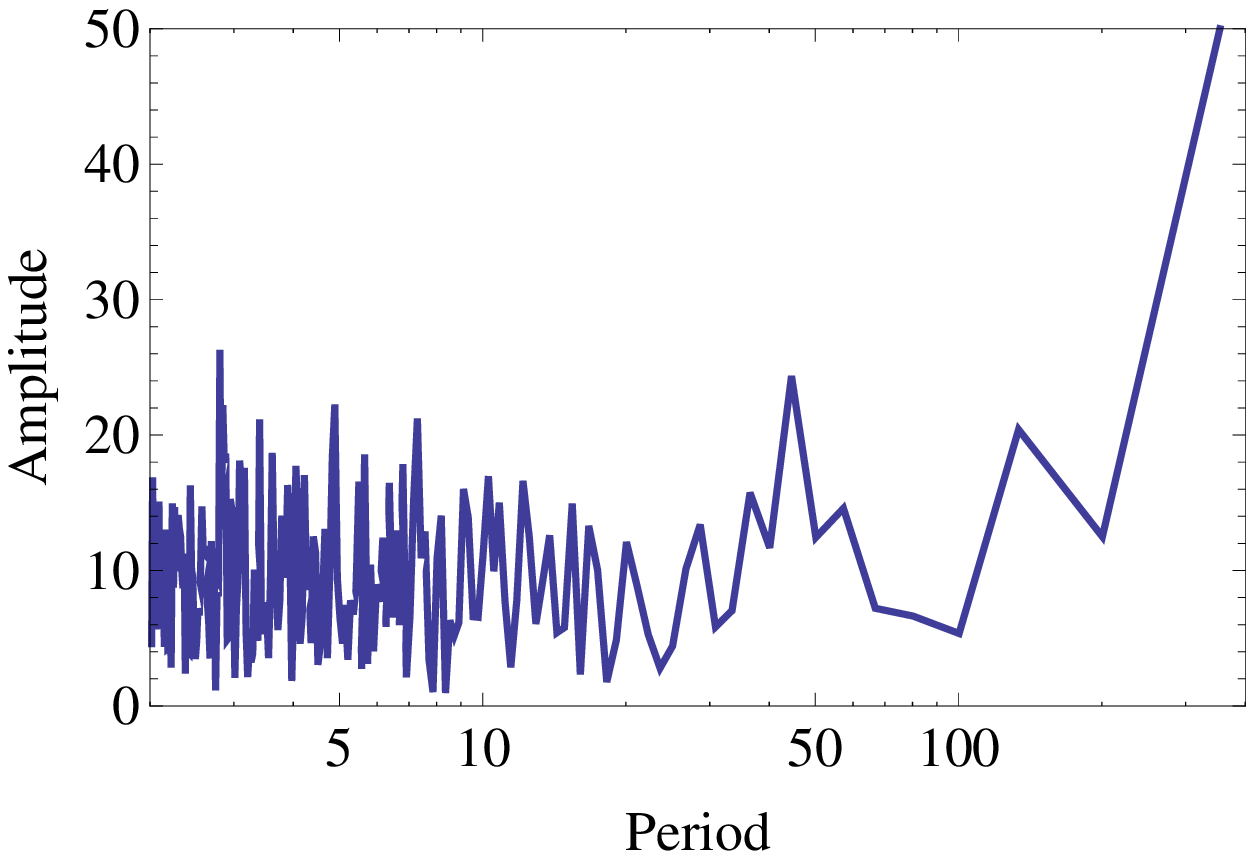}}\\
\subfigure[]{
\includegraphics[scale=0.4]{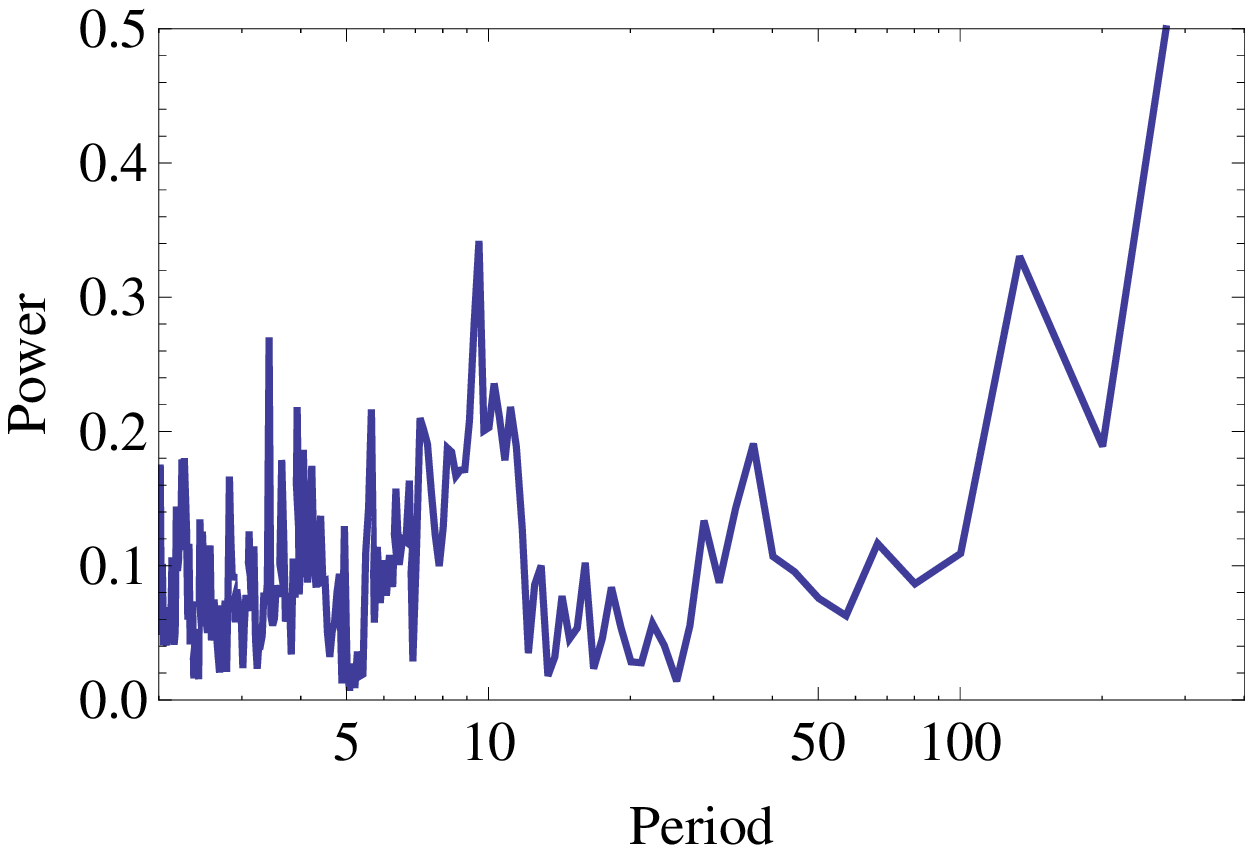}}
\subfigure[]{
\includegraphics[scale=0.4]{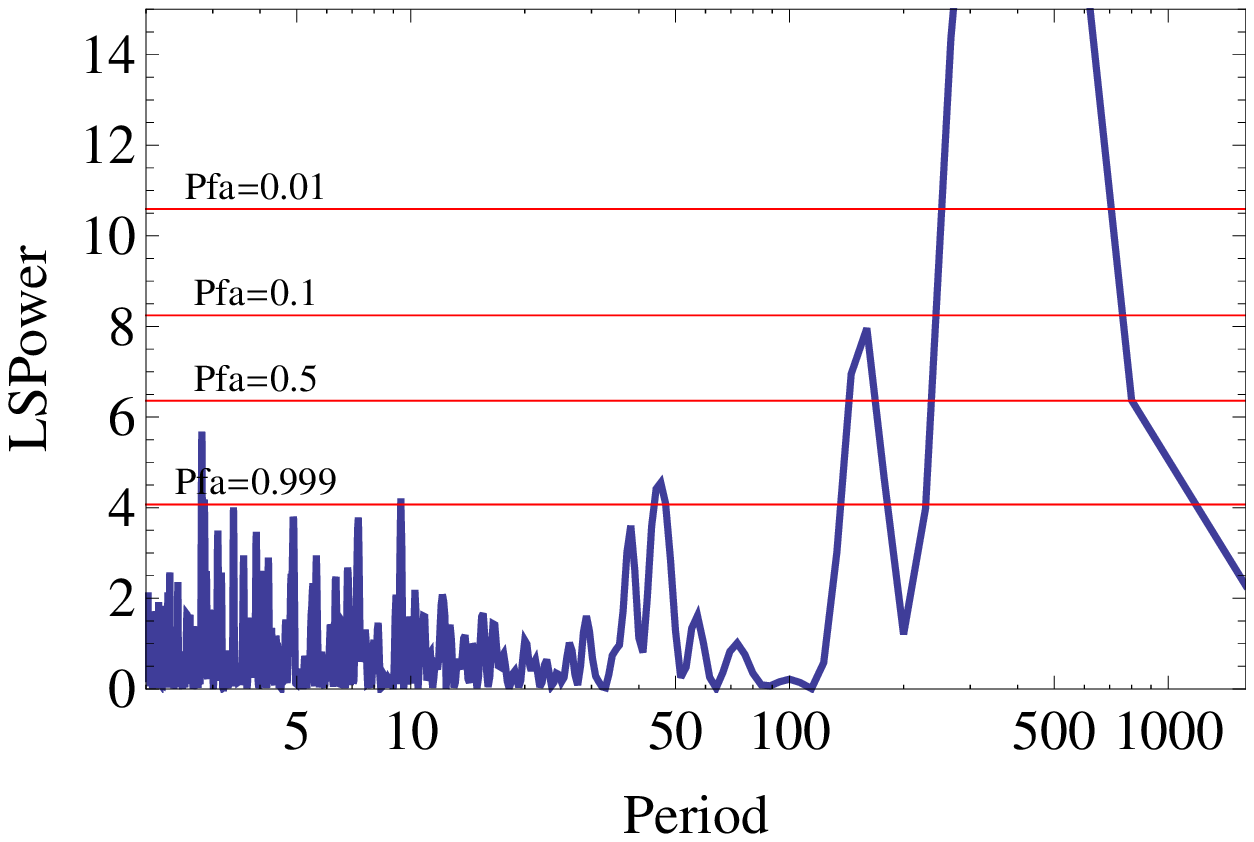}}
\caption{\label{fig:sigUrF}\small (a) A noisy multi-cyclic signal (Eq.~\ref{eq:UrF}), (b) its Fourier spectrum, (c) the power spectrum and (d) the Lomb-Scargle periodgram. In (d) the ``Pfa'' indicates the false-alarm probability.}
\end{center}
\end{figure}

Consider a multi-cyclic signal with noise:
\begin{linenomath}\begin{align}\label{eq:UrF}
u(r)=U_0+U_1\cos\left[\dfrac{2\pi r\Delta t}{T_1}+\phi_1\right]+U_2\cos\left[\dfrac{2\pi r\Delta t}{T_2}+\phi_2\right]++U_3\cos\left[\dfrac{2\pi r\Delta t}{T_3}+\phi_3\right]+w(r),
\end{align}\end{linenomath}
where $U_0=10,\ U_1=5,\ U_2=1,\ U_3=1,\ T_1=420,\ T_2=50,\ T_3=12, \ \phi_1=0.5,\ \phi_2=1,\ \phi=0.7,\ \Delta t=1,$ and $N=800$. The index $r$ runs from $\tfrac{N}{2}$ to $N$ to avoid the initial transition. The first mode corresponds to the ``whole time trend'', the second -- a mid-term cycle and the third -- a short-term cycle. A random signal $w$ has a distribution of $N(0,10)$ that is much stronger than the cycles. This signal and its Fourier and power spectra are shown in Fig.~\ref{fig:sigUrF}. We can see that the second and third modes of $T_2=50,\ T_3=12$ seem to be almost spurious in spectra. Especially, in the Lomb-Scargle periodogram (Fig.~\ref{fig:sigUrF}(d)) the false-alarm probabilities for those cycles are near to 0.999, which means that those cycle might be almost noisy and can not be considered deterministic. 

\begin{figure}
\begin{center}
\subfigure[]{
\includegraphics[scale=0.4]{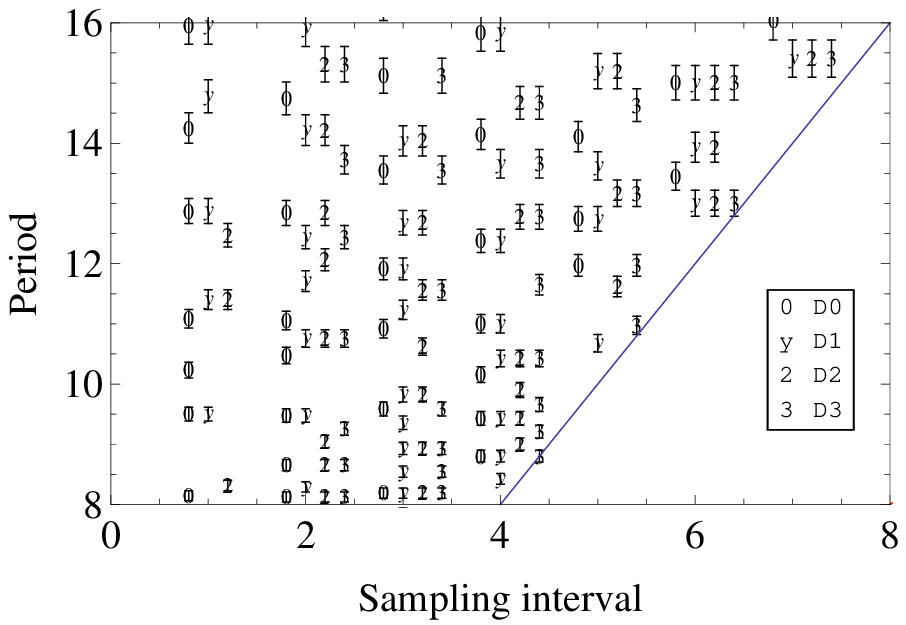}}
\subfigure[]{
\includegraphics[scale=0.4]{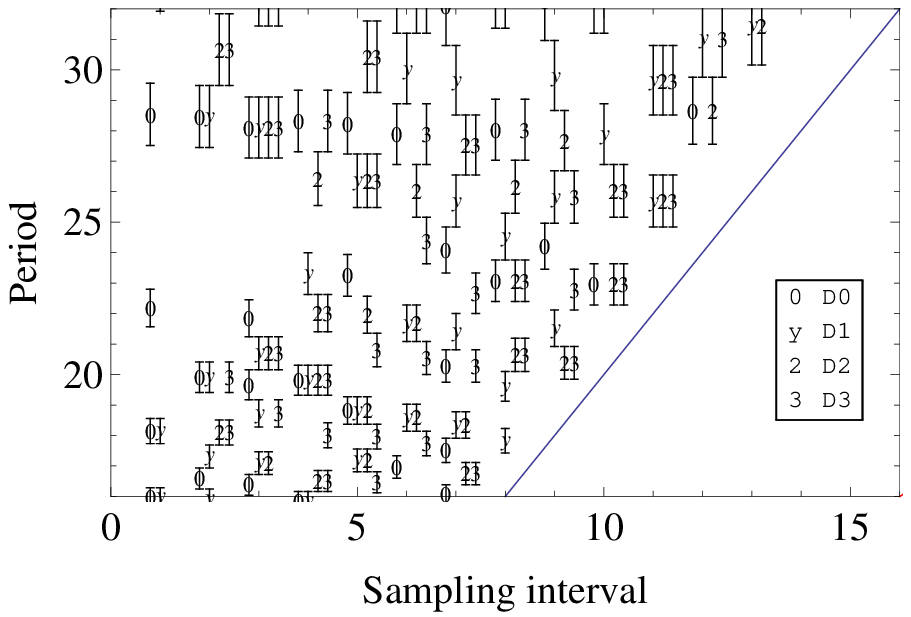}}\\
\subfigure[]{
\includegraphics[scale=0.4]{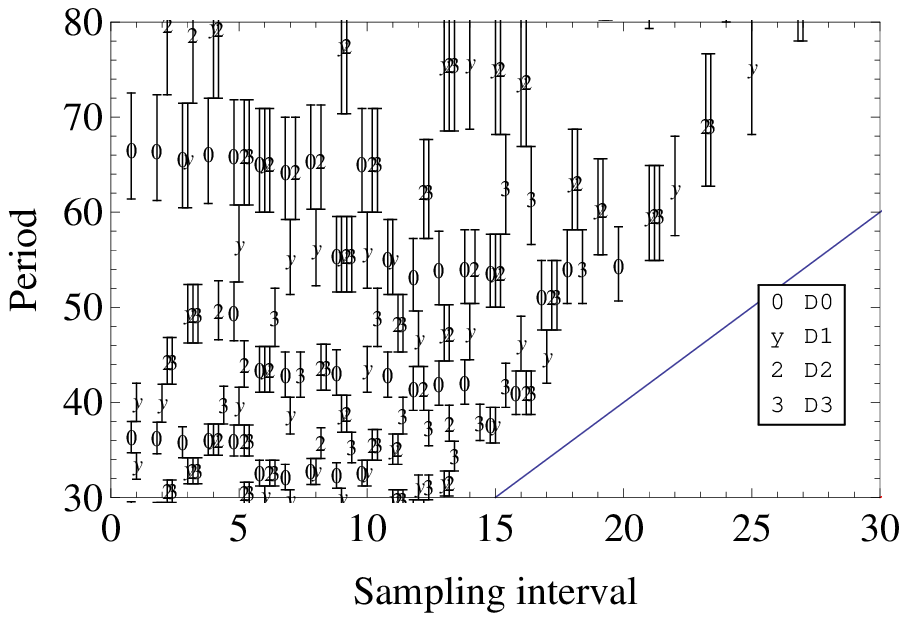}}
\subfigure[]{
\includegraphics[scale=0.4]{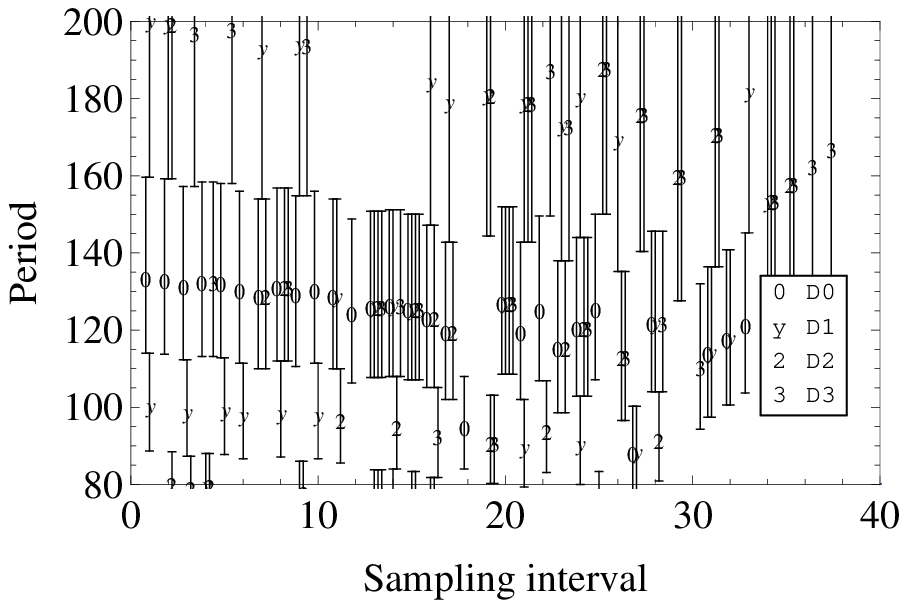}}
\caption{\label{fig:smpUrF}\small The peaks of (a) the samplogram, (b) the D1-samplogram, (c) the D2-samplogram and (d) the D3-samplogram of the noisy multi-cyclic signal (Eq.~\ref{eq:UrF}). The error bar stands for the rounding region of frequency index. The blue diagonal means the Nyquist sampling. The marks of ``0'', ``y'', ``2'', ``3'' mean the peaks of the samplogram and D1 to D3-samplograms.}
\end{center}
\end{figure}

We test the stability of cycles in the samplogram and D-samplograms. Figure~\ref{fig:smpUrF} shows the peaks of the samplogram and D-samplograms up to the 3rd order. We consider that the real decimal frequency is rounded to integer in DFT so we show the rounding region of a given peak within $\pm0.5$ of the frequency index as the error bar (cf. the peak barring in \S~\ref{subsec:lifting}). The peaks are marked with ``0'', ``y'', ``2'' and ``3'' corresponding to the order of D-samplograms. The figure shows only the true periodicity region to the Nyquist sampling.

As we can see, in the short-cycle region (i.e. ``noise'' region, Fig.~\ref{fig:smpUrF}(a)) we can see that the cycles of 9.5, 13 time steps (t.s.) are stable. The high-order peaks appear in some great sampling interval, which means that they are weak. The peak of 66 t.s. is shifting to 55 t.s. when the sampling is increasing. We can estimate also a cycle of $\sim$45 t.s. The weak cycle of 137 t.s. could be supposed. The 9.5 and 137-t.s. cycles are obviously the spurious cycles.
In spite of appearance of the spurious cycles, the example shows that the stochastic stability can filter many other spurious cycles effectively.

The shorter region than 8 t.s. that we neglect here have only 3 or less sampling steps to check the stability. The shorter cycles have the higher resolution of the period. And these cycles almost correspond to ``the noise cycles''. So we would meet the stronger requirement for the shorter cycles that the multiplet of the peaks of various orders should be consistent within the rounding region. However, the longer cycles have the wide variance of the peaks due to low resolution and long range of sampling interval to check the stability. Even the ridge bending appears for the long cycle. The consistency of the peaks of various orders are expected only near the Nyquist sampling. So we can require ``the consistency'' for the short cycle and ``the persistency'' for the long cycle. 

If the cycle is significant, its stability is obvious because in spite of the degradation in the down-sampling the ridge will remain significant. However, the example proves that the non-significant cycles have some stability against the down-sampling and differencing.

\subsection{The spurious cycle and grand extremes in the long-term solar activity}\label{subsec:spurgrand}
Beside manifesting the effectivity of the samplogram analysis, the above example also shows that the features of the sinusoidal stationary cycle would appear in the samplogram even for a random signal. This is because the random signal is decomposed into the sinusoidal signals in the Fourier analysis. However, among those sinusoidal stationary cycles there are the spurious cycles. What is the origin of those spurious cycles? Of course, they come from the random noise. But what we are interested in is what process the significant spurious cycles are made through. If we could know this process, we could find the stable cycles more precisely. We can consider several factors:

First, the spurious cycles can come from the non-sinusoidal waveform. For example, the rectangular or triangular signals cause many spurious cycles. The current SSN dataset accompanies the 11-yr cycle as well as 5.5-yr cycle, which is related to its non-sinusoidal waveform \citep{Petrovay2010}. However, such harmonics sometimes become the essential argument to detect any cycle in astronomical phenomena \citep{Kollath2009, Olah2009} and are not bothering in our case of solar activity. 

Secondly, the spurious cycles can come from the random occurrences of the grand extremes such as grand minima and grand maxima. As we have seen in \S~\ref{subsec:zeropad}, even a short time-segment of cycle, if its length is longer than one period, can make a peak in the spectrum. Therefore a intermittent cycle or even only one separation between the two grand extremes could pretend to be a stationary and even significant cycle. If the similar separation repeats, the peak in spectrum will get higher. Such random separation between the grand extremes is dangerous to the samplogram analysis and any other spectral analyses, though the degradation of the peak in the averaging down-sampling should be much random rather than the normal degradation of the stationary cycle. 

We inspect the separations between the adjacent grand minima and maxima. We use the lists of grand extremes given in \citet{Inceoglu2015} (where we choose the grand minima and maxima reconstructed from $^{10}$Be) and \citet{Usoskin2016}. We plot the histogram for the separations with bin of 20-yr width. The low count (equal to 1) probably means the spurious cycle because this is non-iterative. However, if the count is no less than 2, we can expect the iterative appearances of the separation so this might mean a stable cycle. 

\begin{figure}
\centering
\subfigure[]{
\includegraphics[width=0.2\columnwidth]{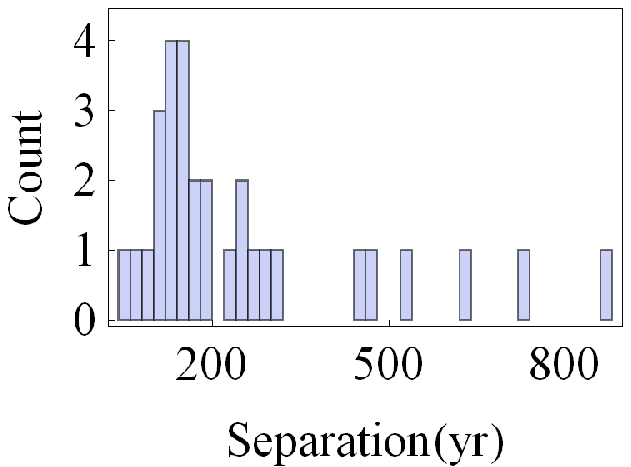}}
\subfigure[]{
\includegraphics[width=0.2\columnwidth]{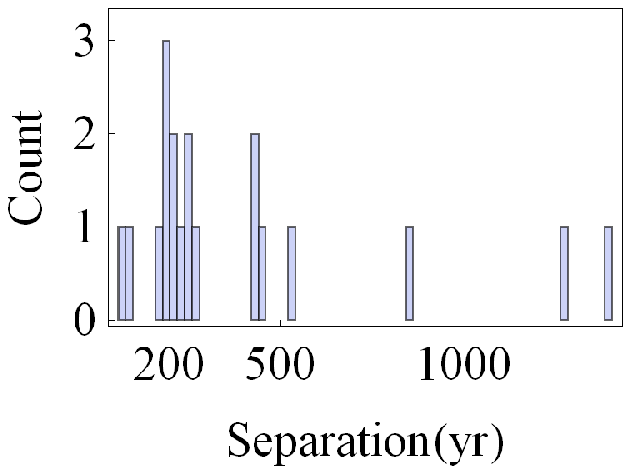}}
\subfigure[]{
\includegraphics[width=0.2\columnwidth]{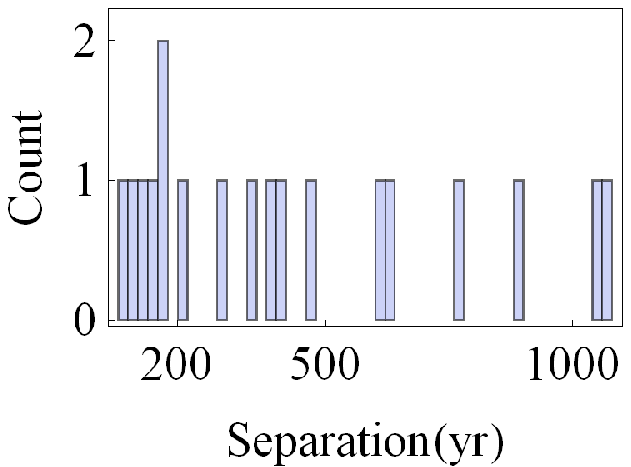}}
\subfigure[]{
\includegraphics[width=0.2\columnwidth]{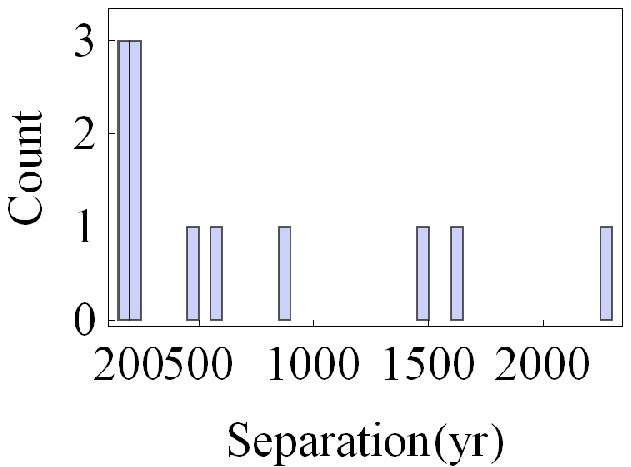}}
 \caption{\label{fig:HisG} The histogram for the separations between grand extremes of reconstructed solar activity. (a) stands for the grand minima and (b) for the maxima from $^{10}$Be reconstruction in \citet{Inceoglu2015}, while (c) for the grand minima and (d) maxima in \citet{Usoskin2016}.}
\end{figure}

We can find the Suess/de Vriece cycle appear more frequently between the grand maxima rather than the grand minima. This is consistent with the result obtained in \citet{Chol-jun2020a} where we verified this result with the superposed epoch analysis (SEA) method. Beside this cycle, we can see the 130, 170, 190, 250-yr separations to be more iterative. 

\subsection{The sampling and differencing stability of cycle}\label{subsec:SnD}

In fact, the power spectrum is based on the Fourier approach where an arbitrary signal is decomposed into the sinusoidal stationary cycles. So the random signal is also transformed into the series of sinusoidal cycles and is often regarded to have short period and weak power. This transformation should be, of course, arbitrary and non-stable in any operation on the signal such as the down-sampling, because the down-sampling includes the averaging in itself. In deed, we can observe that the ridge of spurious cycle degrades much randomly and does not follow the normal degradation, which affects the maintenance of ridge in the averaging down sampling. So we can expect that the spurious cycles representing the random signal should vary in operations such as down-sampling and differencing. 

However, for the deterministic cycle we can assume two properties:
\begin{itemize}
\item [-] the (down-) sampling stability.

and

\item [-] the differencing stability.
\end{itemize}
The (down-) sampling stability implies that the peak for the cycle must be kept when the sampling interval changes (see \S~\ref{subsec:apprps}). This property is proved by the aforementioned ridge in the samplogram. The differencing stablity requires that the peak for the periodicity should be kept in differending (see \S~\ref{subsec:Dlsmplgrm}). This property is checked by matching ridges or peaks of the samplogram and D-samplograms. As said above, the power spectrum of the differences is deformed from that of the original time series. The peaks of longer cycles become lower. However, the peaks are elevated when approaching to the Nyquist sampling. So the matching between the D-samplograms will be more easily attainable near the Nyquist sampling.

For an instance of the traditional spectral analysis, the Lomb-Scargle (LS) periodogram shows a stochastic significance of the cycle by a false-alarm probability (FAP) \citep{Lomb1976, Scargle1989}. This probability indicates how difficult to make the significant peak by a pure noise (Fig.~\ref{fig:sigUrF}). In comparison with this, the sampling and differencing (SnD) stability in samplogram will give another property -- stability of the cycle in a random signal. The significance analysis has a drawback: a weak true cycle can be missed and a spurious but significant cycle can be elected. Even the SnD stability may be affected by the significance because the significant peak has the longer persistence than the weaker one in down-sampling. However, the SnD stability is more effective than the LS periodogram to pick stable cycle, which has been shown in the example of \S~\ref{subsec:prac}. This is because a spurious cycle has less stability (e.g. persistence in the down-sampling) than the true cycle.

As said in \S~\ref{sec:intro}, the drawback of the LS periodgram comes from that peak reflects the frequency consistency and the cycle amplitude. If we want to remove the significant spurious cycle, we should divide both factors. So we introduced the cycle-keeping operations: the down-sampling and differencing. What we want from these operations is the power variation of the cycles. As we have seen in \S~\ref{subsec:powvar}, when the sampling interval increasing, the power of short cycle and long cycle vary differently: the total energy of the signal is concentrated to the long cycle. On the other hand, in \S~\ref{subsec:Dlsmplgrm} we have seen that the total energy is concentrated to the short cycle by differencing. Those trends may be related to making use of power for the normalized autocorrelation (Eq.~\ref{eq:ac1}): if all other cycles but one cycle degrade, then the normalized power of remained cycle will raise, though this cycle does not change. The normalized power implies the energy fraction of signal. Thus the opposite effects of both operations on power can make both operations to compose a set of (at least weakly) independent elementary operations for power variation. Their synergy effect is manifested in the D-samplogram. Around the Nyquist sampling there appears a sequence of the elevated peaks of cycles, even though their resonance peaks might be neglected in spectrum (e.g. engulfed in the adjacent significant peak or background). 

\subsection{The current problems}
So far we have seen the analysis of the stable cycles in the samplogram. However, there are some problems in the analysis.

First, the samplogram is effective to remove the high-frequency spurious cycles but not for the low-frequency ones. This is related to that the averaging suppresses only the high-frequency or short-period noise. We have seen that long spurious cycles remain significant in the samplogram.

Secondly, no quantity is developed still yet to evaluate the stochastic stability of the cycle. We want the quantity like the false-alarm probability to evaluate the stochastic significance. If there is no such a quantity, we have to perform some tedious and ineffective work in terms of examination with the naked eye to match the D-samplograms and analyze the sampling stability.

Thirdly, we can see a ridge bending in the samplogram for some signal. The ridge bending means that the true ridge bends up (to the longer period) or down (to the short period) when the sampling interval is increasing, in particular near the Nyquist sampling. This may be related to the averaging sampling itself or the signal peculiarity. The former may come from that we have truncated the time series. The latter may come from the modulation in signal. The ridge bending makes a trouble to the precise estimation of the period. 

We can consider another way to evaluate the stability of cycle. And we can add a signal artificially. If we add a shortest and significant signal, the all existing cycles including even short cycle will have the ridges elevating around the Nyquist sampling.

We see samplogram analysis for the solar activity in \citet{Chol-jun2020b}.

\section{Conclusion}
We introduced a samplogram analysis that can infer stable cycles by applying cycle-keeping operations to any signal. The spurious cycle is unstable in those operations. In this paper, we considered a averaging down-sampling and differencing as such cycle-keeping operations. In a averaging down-sampling the long cycle is amplified and the short cycle is depressed while in differencing vice versa.

The spectral analysis with stochastic significance cannot distinguish the spurious cycles effectively, because the peak in spectrum involves not only the frequency consistence but also the amplitude of the signal. Even non-frequency-consistent or non-stationary random rise and fall of signal can make a significant peak in spectrum. We showed that the random separation between the grand extremes in long-term solar activity can be misunderstood as a significant stationary (spurious) cycle in traditional spectral analysis.

The sampling and differencing stability can separate the amplitude from the frequency stability in peak height in spectrum, at least weakly, through the amplitude variation in various operations. We saw that this method is more effective than a spectral analysis with stochastic significance in distinguishing the spurious and stable cycles.

\section*{Acknowledgement}

K. Chol-jun has been supported by \text{Kim Il Sung} University during investigation.

\end{document}